\documentstyle[preprint,aps,eqsecnum]{revtex}

\begin{document}

\title {Kac--Moody Symmetries of Critical Ground States}

\author{Jan\'{e} Kondev} \address{Physics Department, Brown
University, Providence, Rhode Island 02912-1843}

\author{Christopher L.  Henley} \address{Laboratory of Atomic and
Solid State Physics, Cornell University, \\ Ithaca, New York, 14853}

\date{\today}

\maketitle

\begin{abstract}

The symmetries of critical ground states of two-dimensional lattice
models are investigated.  We show how mapping a critical ground state
to a model of a rough
interface can be used to identify
the chiral symmetry algebra of the conformal field theory that
describes its scaling limit.
This is demonstrated in the case of the six-vertex model, the
three-coloring model on the honeycomb lattice, and the four-coloring
model on the square lattice. These models are critical and they are
described in the continuum by conformal field theories whose symmetry
algebras are the $su(2)_{k=1}$, $su(3)_{k=1}$, and the $su(4)_{k=1}$
Kac-Moody algebra, respectively.
Our approach is based on the Frenkel--Kac--Segal vertex operator
construction of level one Kac--Moody algebras.

\end{abstract}


\newcommand{\ba}{\mbox{\boldmath $\alpha$}}
\newcommand{\bb}{\mbox{\boldmath $\beta$}}
\newcommand{\bl}{\mbox{\boldmath $\lambda$}}
\newcommand{\br}{\mbox{\boldmath $\rho$}}
\newcommand{\bs}{\mbox{\boldmath $\sigma$}}
\newcommand{\bd}{\mbox{\boldmath $\delta$}}

\newcommand{\be}[1]{\begin{equation}\label{#1}}
\newcommand{\ee}{\end{equation}}
\newcommand{\bea}[1]{\begin{eqnarray}\label{#1}}
\newcommand{\eea}{\end{eqnarray}} \newcommand{\Eq}[1]{Eq.~(\ref{#1})}
\newcommand{\Fig}[1]{Fig.~\ref{#1}} \newcommand{\Rs}{\sf I\hskip-1.5pt
R} \newcommand{\Zs}{\mbox{\sf Z\hskip-5pt Z}} \newcommand{\Cs}{\rm
C\!\!\!I\:}


\section{Introduction}

Over the years the ground state ensembles of certain two dimensional
classical spin models have been found to exhibit critical behavior.
The first such ground state ensemble was encountered in the
antiferromagnetic Ising model on the triangular
lattice~\cite{Cblote,Cnight}.  The correlation functions of operators
constructed from the Ising spins, in the ground state ensemble, were
found to decay with distance as power laws with various
exponents. Typically, critical ground states are found in classical
two-dimensional antiferromagnetic spin models, and in other models
that exhibit frustration.  The existence of critical ground states was
first argued  on rather general grounds, by Berker and Kadanoff 
\cite{Cberker}. 

Other than the Ising antiferromagnet on the triangular lattice, 
known models with critical ground states are the six-vertex
model~\cite{Cliebwu}, the closely related three-state
antiferromagnetic Potts model on the square lattice~\cite{Ckotecky},
the three-state antiferromagnetic Potts model on the Kagom\'{e}
lattice \cite{Cbroholm,Chuse,Cchandra}, 
the four-state antiferromagnetic Potts
vertex model on the square lattice \cite{Cread,CKHPRB}, and the $O(n)$
model on the honeycomb lattice \cite{Cbloteprl,Cbatch,Cmycomm}. New
models that are of this type are the antiferromagnetic Ising model of
general spin on the triangular lattice \cite{Cnagai,Cchenclh}, the
non-crossing dimer model and the dimer-loop model \cite{Carouh}, both
defined on the square lattice.

It has recently been shown that the ground state properties of all the
known classical spin models with critical ground states can be
analyzed by mapping the spin model onto an interface model
\cite{Cclhrev}.  This analysis is equivalent, by a duality, to the
Coulomb gas method, which has been used successfully for calculating
exact values of critical exponents for many two dimensional lattice
models \cite{Cnienrev}. The free energy of the interface is entropic
in origin and the strength of the height fluctuations are governed by
the {\em stiffness} of the interface.  If the interface is in the
rough phase then correlation functions in the spin model decay as
power laws, with exponents whose numerical values are completely
determined by the stiffness.  This scenario is also found in the
Coulomb gas approach to critical phenomena, where the scaling
dimensions of all the electric and magnetic type operators are fixed
by the value of the coupling constant.

Critical ground states are completely defined by constraints on the 
allowed configurations. 
An interesting question that has been raised recently \cite{Cbloteprl}
is how, and whether, the introduction of frustration, or of
constraints, can lead to large values of the
conformal charge ($c>1$) in two dimensional lattice models.  It should
be mentioned that it is a widely held belief that conformal field
theories with large values of the conformal charge have rather limited
application in describing critical phenomena in two dimensions.
Nonetheless, large values of $c$ have been observed in the fully
frustrated XY model on the square lattice \cite{CFFXY}, and more
recently in the fully packed loop model on the honeycomb lattice
\cite{Cbloteprl} and the four-coloring model on the square lattice
\cite{CKHPRB}.  In the loop model, and the coloring model, the large
values of the conformal charge can be understood by mapping these
models to interface models.  The basic idea is that the constraints on
the allowed states in these models require the height, which defines
the interface, to have more than one component.  Therefore the 
interface can be
thought of as an embedding of the two-dimensional lattice in
a higher dimensional target space. If the interface is rough, then
the effective
field theory of the interface, 
which describes the long-wavelength fluctuations of the height, 
is conformal, and its conformal charge is equal to the
dimensionality of the target space; this is well known from the theory
of the bosonic string~\cite{Cgreen}. Furthermore, if the 
target space is {\em compactified} to a torus, then this can give rise to 
a nonabelian symmetry of the effective field theory.

In this paper we analyze the symmetry algebra of the conformal field
theory that describes the scaling limit of the six-vertex model, the
three-coloring model on the honeycomb lattice, and the four-coloring
model on the square lattice, from a unifying perspective.
 This is provided by  a mapping of these
models to models of rough interfaces. The heights that define the
interface are shown to be compactified on the root lattice of
the $su(2)$, $su(3)$, and $su(4)$ Lie algebra, respectively. The {\em
stiffness} of the interface plays the role of the compactification
radius, and for special values of the stiffness a Kac-Moody 
algebra appears in the scaling limit of the interface model.  We
calculate the stiffness {\em exactly} using the loop correlation
function \cite{CKHPRL},
which measures the probability that two points lie on the same contour
loop of the interface; when the height has more than one component
contours of a particular component need to be considered.  Once the
value of the stiffness is known then the scaling dimensions of all the
operators in the six-vertex, the three-coloring, and the four-coloring
model, can be calculated exactly. Furthermore, for each of these
models, the currents of the Kac-Moody symmetry algebra can be written
down in terms of the heigh field using the Frenkel--Kac--Segal
\cite{Cfks} vertex operator construction.

This paper is organized as follows: In Section~\ref{mapping_gen} we
describe the mapping of critical ground states to interface
models. This allows us to introduce the concepts of the ideal state,
the ideal state graph, and the repeat lattice, which play an important
role in the later sections where we study specific models.  In this
section we also introduce an effective field theory for the
long-wavelength fluctuations of the interface, which is conformally
invariant.  The conformal properties of this field theory, and in
particular its chiral symmetry algebra, are discussed in
Section~\ref{conf_gen}.  These first two sections are devoted to
critical ground states in general, and in the remaining sections we
turn to specific examples.  In Section~\ref{6vmodel} we review the
mapping of the six-vertex model to an interface model, we calculate
the stiffness, and we show that this model has a chiral $su(2)_{k=1}$
symmetry. We follow the same procedure in Sections~\ref{3colmod}
and~\ref{4colmod}, where we show that the the scaling limits of the
three-coloring, and the four coloring model are the $su(3)_{k=1}$, and
the $su(4)_{k=1}$ Wess--Zumino--Witten model \cite{Cwitt}, respectively.

\section{Interface representations of critical ground states}
\label{mapping_gen}

Here we give a brief summary of the construction of an interface model
equivalent to the critical ground state of a spin model; the details
of this construction, for the specific models, will be given in
Sections~\ref{6v_interface}, \ref{3col-interface},
and~\ref{4col-interface}. First we discuss the mapping itself which,
after coarse graining, leads to an effective field theory for the
long-wavelength fluctuations of the interface. In the second part of
this section we show how operators in the ground state ensemble can be
expressed in terms of the height field, and how their scaling
dimensions can be calculated once the stiffness of the interface is
known.  In the third part we outline the calculation of the stiffness,
which relies on the so-called {\em loop correlation function}. This
correlation function measures the probability that two points on a
rough interface lie on the same contour loop \cite{CKHPRL}.

\subsection{Mapping to an interface model and the effective field theory}
\label{eff_gen}

Critical ground states are defined by the constraints on the allowed
spin configurations. For example, in every ground state of the
antiferromagnetic Ising model on the triangular lattice, all
triangular plaquettes have exactly one pair of spins pointing in the
same direction, or in other words, one pair of ``frustrated'' spins.
Every ground state is given the same statistical weight.

In order to map a critical ground state of a spin model onto an
interface model we start by defining a {\em height rule}.  The height
rule allows us to map a given ground state spin configuration $\{ {\bf
\sigma}({\bf x}_i) \}$ to a {\em microscopic height} configuration $\{
{\bf z}({\bf x}_i) \}$.  In general, the microscopic heights take
their values in $\Zs^D$, and they define a two-dimensional interface
in $D+2$ dimensions. Each allowed height configuration of the
interface is given the {\em same} statistical weight.

Typically the height rule is such that the difference in ${\bf z}$
between neighboring sites ${\bf x}_i$ and ${\bf x}_j$ on the lattice,
is determined by the spins $\sigma({\bf x}_i)$ and $\sigma({\bf
x}_j)$.
In order for the height rule to be well defined, the change in the
microscopic height must be zero when a closed loop on the lattice is
traversed.  Furthermore, we consider a mapping given by the height
rule to be a {\em height mapping} if one can always find a spin
configuration that will have any given average slope of the
microscopic height; this ensures that ${\bf z}$ can take any value in 
$\Zs^D$, i.e., the height is not restricted. 

In order to define an effective field theory for the interface model,
we introduce a coarse-graining procedure for the microscopic heights;
see Fig.~\ref{Csketch} for a summary.  First, among the ground states
of the spin model we identify {\em ideal states} which are {\em flat},
that is they have zero average slope and they minimize deviations of
the microscopic height away from its average.
These ground states also have a maximum entropy density, that is they
allow for a maximum number of local rearrangements of the spins which
are in accord with 
the ground state constraints.  The implicit assumption being
made is that states with the maximum entropy density are flat. We will
find this to be the case for all the models studied here.

Second, we coarse-grain  the original spin model using the ideal states.
We think of the lattice as being broken up 
into {\em ideal state domains};
in each domain fluctuations, which simply correspond to rearrangements
of the spins allowed by the ground state constraints,
occur about a particular ideal state.
On the level of the interface, the microscopic heights ${\bf z}$ are
replaced by the {\em coarse grained heights} ${\bf h}$, which are
defined for every ideal state domain, and equal to the average
microscopic height in a particular domain (see Fig.~\ref{Csketch}(b)):
${\bf h}=<{\bf z}>$.

Third, we define the {\em ideal state graph} ${\cal I} \subset
\Rs^D$. Every node of ${\cal I}$ represents an ideal state, and its
position in $\Rs^D$ is given by the coarse grained height ${\bf h} \in
\Rs^D$ of the ideal state it represents. Two nodes of the graph ${\cal
I}$ are connected if the two ideal states they represent differ by the
minimum number of local spin rearrangements needed to transform one
ideal state to the other.

Two different points on the ideal-state graph ${\bf h}$ and ${\bf h}'$
can correspond to the same spin configuration (ideal state). The set
of vectors ${\bf b} = {\bf h} - {\bf h}'$ form a lattice in $\Rs^D$
called the {\em repeat lattice} ${\cal R}$. This lattice plays an
important role in calculating scaling dimensions of lattice operators,
which are defined as local functions of the spins, and it holds the
key to identifying the symmetry algebra of the scaling limit of the
critical ground states.
 
Finally, we consider the continuum limit of the interface model, where
the heights defined over particular ideal state domains are replaced
with a continuously varying height field ${\bf h}({\bf x}) \equiv
(h_1({\bf x}), h_2({\bf x}),\ldots, h_D({\bf x}))$, as seen in
Fig.~\ref{Csketch}(c).  The dimensionless free energy (action) of the
interface, which is entropic in origin, is assumed to be of the
form 
\begin{equation}
\label{free}
f = \int d^2 {\bf x} \: \left[(\frac{K}{2} \sum_{i=1}^D |\nabla h_i({
\bf x})|^2 + V({\bf h}) \right] \: ;
\end{equation}
$K$ is the {\em stiffness} of the interface,\footnote{Here we 
assume that the stiffness is an isotropic
tensor, which is not always the case. Critical ground states with more
complicated stiffness tensors have been defined and analyzed in
\cite{Carouh}.} and
$V({\bf h})$ is a periodic potential with the periodicity given
by the ideal state graph, i.e., 
\begin{equation}
\label{periodicity1}
     V({\bf h} + {\cal I}) = V({\bf h}) \; .
\end{equation}

The free energy $f$ defines an effective field theory of the critical
ground state; the assumption being made is that it correctly describes
the long-wavelength fluctuations of the microscopic height ${\bf z}$.
The periodic potential $V({\bf h})$, which is usually referred to as
the {\em locking potential}~\cite{CJKKN}, favors the heights to take
their values on ${\cal I}$, while the first term represents
fluctuations around the flat ideal states.  Therefore, the assumption
that the effective field theory of the critical ground state is given
by Eq.~(\ref{free}), is directly related to the intuitive idea put
forward earlier, that the free energy (entropy) of the ground state is
entirely due to fluctuations around the ideal states.

The locking potential is periodic with the periodicity of ${\cal I}$.
Thus, the ground state, in its interface representation, undergoes a
{\em roughening transition} for some value of the stiffness $K =
K_r$~\cite{CJKKN}. If the stiffness $K$ satisfies $K<K_r$, then $
V({\bf h})$ in Eq.~(\ref{free}) is irrelevant, in the renormalization
group sense, the ground state is critical,
and its scaling limit is
described by a Gaussian model, with the free energy
\begin{equation}
\label{freegauss}
 f = \frac{K}{2} \int d^2 {\bf x} \: ( \sum_{i=1}^D |\nabla h_i({ \bf
x})|^2 ) \; .
\end{equation}
In the case that the locking term is relevant ($K>K_r$), the ground
state will lock into long range order in one of the ideal states.\footnote
{Note that our
analysis implies that a ground state ensemble, for which a height
mapping can be defined, is either critical (i.e. the interface is
rough) or long range ordered (i.e. the interface is smooth); in
particular, it can {\em not} be disordered.}

\subsection{Operators and critical exponents}
\label{ops_gen}

Here we discuss the calculation, in the ground state ensemble, of two
point correlation functions of lattice operators.  Our analysis is
equivalent to the Coulomb gas approach \cite{Cnienrev}, only now, as
will be shown below, the magnetic and electric charges are {\em
vectors} in the repeat lattice and its reciprocal.
Coulomb gas methods with vector charges have been used previously by
Fateev and Zamolodchikov \cite{Cfatzam} to calculate correlation
functions in the $\Zs_3$ models. Their work was extended by Pasquier
\cite{Cpasq} who considered the continuum limit of lattice models with
quantum group symmetries.

A local operator $O({\bf x})$ constructed from the spins and spatially
uniform in the ideal states, can be written in terms of the coarse
grained height as $\overline{O}({\bf h}({\bf x}))$. Since ${\bf h}$
and ${\bf h}+{\cal R}$ represent the same ideal state on the ideal
state graph, we can identify them
\begin{equation}
\label{compactify1}
{\bf h} \equiv \: {\bf h} + {\cal R} \; ,
\end{equation} 
and the operator $\overline{O}({\bf h})$ is necessarily periodic with
the periods forming the repeat lattice ${\cal R}$.  Therefore, we can
write $O({\bf x})$ as a Fourier series
\begin{equation}
\label{fourier}
     O({\bf x}) = \sum_{{\bf G} \in \cal{R}^\ast} O_{{\bf G}} e^{i
     {\bf G} \cdot {\bf h}({\bf x})}
\end{equation}
where $\cal{R}^\ast$ is the lattice {\em reciprocal} to the repeat
lattice.\footnote{By definition, the dot product of any vector in the
repeat lattice with any vector in the reciprocal lattice is an integer
multiple of $2\pi$.}  The scaling dimension of $O({\bf x})$ is equal
to the scaling dimension of the most relevant {\em vertex operator}
$\exp(i {\bf G} \cdot {\bf h}({\bf x}))$ in the above expansion.  In
this sense, lattice operators constructed from the spins can be
associated with vertex operators in the Gaussian model.

{}From the assumed form of the free energy, Eq.~(\ref{free}), we can
calculate the height-height correlation function:
\begin{equation}
\label{corr}
  < (h_i(0) - h_j({\bf x}))^2 > = \frac{\delta_{ij}}{\pi K} \ln|{\bf
   x}| + {\rm const} \: ,
\end{equation}
for $|{\bf x}| \gg a$, the lattice spacing.  Using Eq.~(\ref{corr}) we
can calculate the two-point correlation function
\begin{equation}
\label{corr2}
    < e^{i{\bf G} \cdot {\bf h}(0)} \: e^{-i{\bf G} \cdot {\bf h}({\bf
   x})} > \: = e^{-\frac{1}{2}<[{\bf G} \cdot ({\bf h}(0) - {\bf
   h}({\bf x}))]^2>} \: \sim \frac{1}{|{\bf x}|^{2 x({\bf G})}} \: ,
\end{equation} 
where the exponent $x({\bf G})$ is the scaling dimension of the vertex
operator $\exp(i {\bf G} \cdot {\bf h}({\bf x}))$, and it is related
to the stiffness by
\begin{equation}
\label{dim}
     x({\bf G}) = \frac{1}{2 \pi K} \: \frac{{\bf G}^2}{2} \: .
\end{equation}

The other type of operators we consider correspond to violations of
the ground state condition that is placed on the spins.  This
typically corresponds to a {\em topological defect} (vortex) in the
height language, where the topological charges (Burgers vectors) ${\bf
b}$ take their values in the repeat lattice.\footnote{In the language
of conformal field theory these would be the winding modes, while the
vertex operators are related to the momentum modes \cite{Cginsparg}.}
Using the equation for the free energy, Eq.~(\ref{free}), we find that
the vortex-antivortex correlation function decays algebraically
\cite{Cnienrev}, with the exponent:
\begin{equation}
\label{dimv}
     x_v({\bf b}) = \frac{K}{2 \pi} \: \frac{{\bf b}^2}{2} \: .
\end{equation}
This result follows from the logarithmic form of the dimensionless
interaction energy between two vortices ($\pm{\bf b}$), separated by
$r$: $E_{\rm int}=\frac{K}{2\pi} {\bf b}^2 \ln{r} + {\rm const}$; the
vortex-antivortex correlation function is given by the Boltzmann
factor $\exp{(-E_{\rm int})}$.

We see that all the exponents describing critical correlations in the
ground state of the spin model can be related to the stiffness $K$ of
the interface. Here the stiffness plays the role of the coupling
constant in the Coulomb gas approach \cite{Cnienrev}, or the radius of
compactification of the appropriate conformal field theory
\cite{Cginsparg}.

\subsection{Contour loops}
\label{contours_gen}

We have argued recently \cite{CKHPRL} that geometrical exponents
associated with contour loops on a Gaussian interface have {\em
universal} values which are independent of the stiffness. In
particular, the loop correlation function ${\cal G}({\bf x})$, which
measures the probability that two points on the interface, separated
by ${\bf x}$, belong to the same contour loop, falls off with distance
as 
\be{loopcor} 
{\cal G}({\bf x}) \sim \frac{1}{{\bf |x|}^{2x_{\rm loop}}} \; , 
\ee 
where the loop correlation-function exponent
\be{loopexp} x_{\rm loop} = \frac{1}{2} \: , \ee is independent of the
stiffness.  Therefore, if we could somehow calculate the loop
correlation function in a particular spin model in terms of the
stiffness, then the above result would clearly allow us to find the
exact value of the stiffness. For the critical ground states studied
in this paper this is indeed the case. It will be shown for the
six-vertex model (for a particular choice of vertex weights), the
three-coloring model, and the four-coloring model, that the loop
correlation function is equal to a vortex-antivortex correlation
function for a particular value of the Burgers charge ${\bf b}_{\rm
loop}$.  As was discussed in the previous section the
vortex-antivortex correlation function falls of with distance as a
power law, with the exponent $2x_v({\bf b}_{\rm loop})$ which depends
on the stiffness $K$; see \Eq{dimv}. From \Eq{loopexp} $x_v({\bf
b}_{\rm loop})=1/2$ follows, and the stiffness can be calculated.

\section{Conformal invariance of critical ground states}
\label{conf_gen}

Here we discuss the conformal properties of the field theory which
describes the continuum limit of a critical ground state. In the
previous section we found, under certain assumptions, this to be a
simple Gaussian field theory, and we therefore first give a lightning
review of conformal invariance in the Gaussian model, mostly following
Ginsparg \cite{Cginsparg}.  We also found that the Gaussian fields,
namely the heights, are compactified on the repeat lattice, which
naturally leads into the discussion of the vertex operator (VO)
construction of affine Lie algebras.  We briefly review this
construction for $su(N)_{k=1}$ Kac-Moody (KM) algebras in the second
part of this section; details of the VO construction can be found in
the original papers by Frenkel, Kac, and Segal \cite{Cfks}, and in the
review by Goddard and Olive \cite{Colive}.

\subsection{Gaussian model}

The Gaussian field theory described by \Eq{freegauss} is conformally
invariant; the free energy $f$ can be thought of as the Euclidean
action for $D$ free massless bosons \cite{Cginsparg}. Consequently,
the conformal charge is $c=D$.

If instead of the Cartesian coordinates ${\bf x}=(x_1,x_2)$, we
introduce complex coordinates in the plane, $w=x_1+ix_2$ and
$\bar{w}=x_1-ix_2$, the free energy in \Eq{freegauss} can be written
as: 
\be{B:action} f = K \int d^2 \! w \: \partial{\bf h}(w,\bar{w})
\cdot \bar{\partial} {\bf h}(w,\bar{w}) \; , \ee 
where $\partial=1/2(\partial_1 + \partial_2)$,
$\bar{\partial}=1/2i(\partial_1 - \partial_2)$, and $d^2w = 2d^2{\bf
x}$ due to the Jacobian factor. Written in this form the invariance of
the free energy under a conformal mapping $w'=f(w),
\bar{w}'=\bar{f}(\bar{w})$ is apparent, if we postulate that the
height field transforms as a scalar quantity, i.e.,  \be{B:scalar} {\bf
h}'(w',\bar{w}') = {\bf h}(w,\bar{w}) \; .  \ee

The Euler--Lagrange equation for the height field, with the Euclidean
action given by \Eq{B:action}, is $\bar{\partial}\partial {\bf
h}(w,\bar{w})=0$. We can therefore write the height field as a sum of
a holomorphic and an antiholomorphic field, \be{B:holo} {\bf h}({\bf
x}) \equiv {\bf h}(w,\bar{w}) = {\bf h}(w) + 
\bar{{\bf h}} (\bar{w}) \;.  \ee

Now the holomorphic and antiholomorphic components of the stress-energy
tensor can be expressed in terms of ${\bf h}(w)$ and $\bar{{\bf h}}
(\bar{w})$ as: \be{stresstensor} T(w) = -2 \pi K \: :\partial {\bf
h}(w) \cdot \partial {\bf h}(w): \; \; \; \; \bar{T}(\bar{w}) = - 2
\pi K \: :\bar{\partial} \bar{{\bf h}}(\bar{w}) \cdot \bar{\partial}
\bar{{\bf h}}(\bar{w}): \; , \ee where the symbol $:\ldots:$ denotes
the normal ordering prescription defined by the point splitting
operation \cite{Cginsparg}.  The modes of $T(w)$, \be{B:virgen} L_m =
\oint_C \frac{dw}{2\pi i} w^{m+1} T(w) \; , \ee
satisfy the commutation relations of 
the Virasoro algebra with central charge $c=D$
\be{B:virasoro} [L_m,L_n] = (m-n) L_{m+n} + \frac{D}{12} (m^3 - m)
\delta_{m+n,0} \; , \ee which is the symmetry algebra of the CFT
defined by \Eq{B:action}; the contour integral in \Eq{B:virgen} is
taken around the origin.

The symmetry algebra of the Gaussian field theory, \Eq{B:action}, which
describes a critical ground state, can be larger than Virasoro, which
will necessarily be present as a subalgebra.  This will be the case if
the repeat lattice is equivalent, up to an overall scale factor,
to the {\em root lattice} of some simple Lie algebra $\overline{g}$,
{\em and} the stiffness (compactification radius) has the correct value.
Namely, in this case the symmetry algebra associated with the scaling
limit of a critical ground state is the chiral Kac--Moody algebra $g$
at level one, whose horizontal subalgebra is $\overline{g}$
\cite{Cfuchs}. The currents of $g$ can be constructed from the height
field using the Frankel--Kac--Segal vertex operator construction
\cite{Cfks}, which we review next.

\subsection{Vertex operator construction}

For the purposes of this section we will assume that ${\cal R}\subset
\Rs^D$ is equivalent, up to a scale factor, to the root lattice of an
$su(D+1)$ Lie algebra; the $D(D+1)/2$ {\em positive} roots form a $D$
simplex: equilateral triangle for $D=2$, tetrahedron for $D=3$, and so
on.  We show, for a particular value of the stiffness $K$, that the
symmetry algebra of the Gaussian field theory, \Eq{B:action}, where
the height field is compactified on ${\cal R}$, is the $su(D+1)_{k=1}$
Kac-Moody algebra.

The $su(D+1)_{k=1}$ Kac--Moody (KM) algebra is an infinite dimensional
Lie algebra, whose generators $\{ T^{a}_m : \: a=1,\ldots,(D+1)^2-1\:
; \; m \in \Zs \}$ satisfy the commutation relations \be{B:KM}
[T^{a}_m, T^b_n] = f^{ab}_c T^c_{m+n} + m \delta_{m+n,0} \; , \ee
where $f^{ab}_c$ are the structure functions of the $su(D+1)$ Lie
algebra.  The defining commutation relations of an $su(D+1)_{k=1}$ KM
algebra can be conveniently rewritten in the so-called Cartan--Weyl
basis as \bea{B:KMChevalley} [H^i_m, H^j_n] & = & m \delta^{i,j}
\delta_{m+n,0} \\ \nonumber [H^i_m, E^{\ba}_n] & = & \alpha^i
E^{\ba}_{m+n} \\ \nonumber [E^{\ba}_m, E^{\bb}_n] & = &
\varepsilon(\ba,\bb) E^{\ba + \bb}_{m+n} \; , \; \; \; {\rm for} \; \ba + \bb
\; {\rm a} \; {\rm root} \\ \nonumber [E^{\ba}_n, E^{-\ba}_{m}] & = &
\frac{2}{\ba^2} ( \ba\cdot {\bf H}_{m+n} + n \delta_{m+n,0}) \eea
where $H^i_m$ are the Cartan generators, while $E^{\ba}_m$ are the
step operators associated with the roots ($\ba$) of the $su(D+1)$ Lie
algebra; the constants $\varepsilon(\ba,\bb)$ are antisymmetric in
$\ba$ and $\bb$, and can be normalized to $\pm 1$.  The number of
Cartan generators for an $su(D+1)$ Lie algebra is $D$ while the number
of roots is $D(D+1)$; this gives a total of $(D+1)^2-1$ generators
which is the dimension of $su(D+1)$.

It can be shown that the commutation relations, in the Cartan-Weyl
basis, of an $su(D+1)_{k=1}$ KM algebra are also the commutation
relations of the modes of the {\em currents}
\bea{B:gens} H^i(w) & = & i \partial h^i(w) \\ \nonumber E^{\ba}(w) &
= & \: :e^{i \ba \cdot {\bf h}(w)}: c_{\ba} \; , \eea
where $c_{\ba}$ are the cocycle factors which are needed to give the
proper signs in the commutation relations of the current modes (for
details see Sec. 6.5 in \cite{Colive}).

The derivatives of the height field are currents of the Gaussian field
theory given by \Eq{B:action}\footnote{They are generators of $D$
copies of the $u(1)_{k=1}$ Kac-Moody algebra, which is defined by the
first relation in \Eq{B:KMChevalley}.}  {\em regardless} of the
stiffness $K$, while for this to be true of the {\em vertex}
operators $E^{\ba}(w)$ it is required that their conformal dimensions
$(h,\bar{h})$ satisfy \be{confdim} (\frac{1}{2\pi K} \:
\frac{\ba^2}{4}, 0) = (1, 0) \; .  \ee
We see that the value of the stiffness has to be fine tuned if the
above equation is to hold, since the vectors $\ba$ are completely
determined by the height construction.  Namely, the roots $\ba$ 
generate the root lattice, and they  are also
vectors in ${\cal R}^\ast$, which is to be identified with the {\em
weight} lattice of an $su(D+1)$ Lie algebra. The quotient of the
weight lattice with the root lattice is $\Zs_{D+1}$, which completely
specifies the $D(D+1)$ reciprocal vectors $\ba$, once ${\cal R}^\ast$
is known.

To conclude, two ingredients are necessary, and sufficient, for the
Gaussian field theory, which describes the continuum limit of a
critical ground state, to have a chiral symmetry given by the
$su(D+1)_{k=1}$ Kac-Moody algebra:
1) the repeat lattice has to be equivalent, up to a scale factor, to
the root lattice of the $su(D+1)$ Lie algebra, and 2) the stiffness
has to be such that the vertex operators associated with the root
vectors are dimension-one currents. In the remaining sections of this
paper we
show that both ingredients are to be found in the six-vertex model,
the three-coloring, and the four-coloring model.  Finally, we note
that the stress-energy tensor in \Eq{stresstensor} is in the Sugawara
form, which means that the Gaussian field theory described here is
nothing but the $SU(D+1)_{k=1}$ Wess-Zumino-Witten
model.\footnote{Here, and throughout this paper, we adopt the
definition of a Wess-Zumino-Witten model as a conformal field theory
with a chiral Kac--Moody symmetry algebra and a stress-energy tensor
of the Sugawara form~\cite{Cfuchs}.}

\section{Six-vertex model}
\label{6vmodel}

In this section it is shown that the symmetry algebra of the scaling limit
of the six-vertex model, for a particular choice of vertex weights, is
the $su(2)_{k=1}$ Kac--Moody algebra. Namely,  the effective
field theory of the six vertex model turns out to be 
a Gaussian field theory with a
one-component height compactified on the root lattice of the $su(2)$
Lie algebra.

\subsection{Definitions}
\label{Fdefs}

The six-vertex model on the square lattice is defined by placing
arrows along the bonds of the lattice, with the {\em constraint} that
at each vertex the number of arrows pointing in, is equal to the number
of arrows pointing out. This constraint is the so called ice rule
which has its origins in the charge neutrality condition for square
ice, for which this model was originally proposed~\cite{Cliebwu}.
Namely, in square ice, we think of the oxygen atoms as sitting at the
vertices and the hydrogen atoms on the bonds of a square lattice. Each
arrow points from the hydrogen to the oxygen atom to which it is
bonded, and local 
charge neutrality requires that the divergence of arrows
at each vertex be zero.

There are six possible vertex configurations and to each one we assign
a Boltzmann weight, $a$, $b$, or $c$, as shown in
Fig.~\ref{C6vweights}.  This model was solved exactly by Lieb
\cite{Clieb67}. Recently
Affleck \cite{Caff6v} has shown that the six-vertex model with
isotropic vertex weights, $a=1$, $b=\lambda$, and $c=1+\lambda$, has a
chiral $su(2)_{k=1}$ Kac--Moody symmetry.
This symmetry is hidden, and it can be uncovered by mapping the
six-vertex model to a quantum spin chain \cite{Caff6v}. Here we
demonstrate the same result for the symmetry algebra of the six-vertex
model with vertex weights $a=b=c/2=1$, by mapping it to an interface
model. We mainly do this as an example of the general construction
outlined in the previous section, and to set the stage for the
sections to come.

\subsection{The interface model}
\label{6v_interface}

The six-vertex model can be mapped to an interface model, which is the
so called body-centered solid-on-solid model (BCSOS) introduced by van
Beijeren~\cite{Cvanb}. It describes the surface of a body centered
cubic structure when viewed in the $[100]$ direction.

The microscopic heights of the BCSOS model z $\in \Zs$, are defined at
the centers of the square plaquettes in such a way that when crossing
a bond of the square lattice, z is increased (decreased) by $1$,
depending on whether we cross from left to right (right to left), as
seen when looking in the direction of the arrow on the bond; see
\Fig{C6v_ideal}.  The ice rule ensures that the microscopic heights
are well defined, i.e.,  the total change in height when going around a
closed loop is zero.  In other words, the zero divergence property of
the arrows in the six-vertex model becomes the zero curl property of
the height increments in the BCSOS model.

The smallest change on the lattice, that does not violate the ice rule,
is accomplished by reversing the direction of the arrows on all the
bonds along a loop of constant arrow direction. This observation
prompts us to define ideal states of the six-vertex model as ones that
maximize the number of such loops. There are two ideal states related
to each other by translations of the lattice, and both are defined by
a c-type vertex state which is periodically repeated throughout the
lattice; see Fig.~\ref{C6v_ideal}.
The ideal states also have the important property that they are flat,
in the sense that they minimize the variance of the microscopic
height, and we use them to coarse grain the six-vertex model.

The ideal state graph ${\cal I}$, is a one-dimensional lattice with
the lattice spacing equal to $1$. Each vertex corresponds to one of
the two ideal states, and its position in the graph is determined by
the average microscopic height in the ideal state it represents.  The
same ideal state repeats every two vertices on ${\cal I}$, and these
vertices form the repeat lattice ${\cal R}$, which has a lattice
spacing of $2$; see Fig.~\ref{C6v_ideal_graph}.  The ideal state graph
and the repeat lattice play a crucial role in constructing an
effective field theory of the BCSOS model, which is what we turn to
next.

In order to describe the long-wavelength fluctuations of the
microscopic height in the BCSOS model we introduce a coarse-graining
procedure.  Namely, we think of dividing up the square lattice into
ideal state domains, and
to each of these domains we assign a coarse grained height $h$, which
is equal to the microscopic height averaged over the domain.
The effective field theory for the long-wavelength fluctuations of the
BCSOS model is given by the dimensionless free energy (action), which
is assumed to be of the form: \be{6vfree} f = \int d^2{\bf x} \:
[\frac{K}{2}(\nabla h)^2 + V(h)] \; .  \ee The height field $h({\bf
x})$ is the continuum limit of the discrete coarse grained heights
defined over ideal state domains, and $K$ is the dimensionless
stiffness of the interface.  Since $h$ and $h+{\cal R}$ correspond to
the same ideal state, we consider the height field compactified on the
circle, i.e.,  \be{Fcom} h \in \Rs / {\cal R} \; .  \ee

The potential term $V(h)$ favors heights on the ideal state graph and
is therefore a periodic function of the height field, \be{FVh} V(h +
{\cal I}) = V(h) \; .  \ee The scaling dimension of the periodic
potential is given by \Eq{dim} \be{xV} x({\rm G}_V) =  
\frac{\pi}{K} \; , \ee where 
${\rm G}_V=2 \pi$
is the periodicity of the ideal state graph; see
\Fig{C6v_ideal_graph}.  Therefore, adding the potential term to the
Gaussian free energy is an irrelevant perturbation,
i.e., $x({\rm G}_V)>2$, for \be{FKr} K < \frac{\pi}{2} \; .  \ee
If this condition is satisfied then the effective field theory of the
BCSOS model is purely Gaussian, defined by a dimensionless free energy
\be{6vfree2} f = \frac{K}{2} \int d^2{\bf x} \: (\nabla h)^2 \; .  \ee
If on the other hand $K > \frac{\pi}{2}$, the periodic potential is
relevant, and the BCSOS model will be in the smooth phase, i.e. the
six-vertex model will lock into long-range order in one of the ideal
states.

\subsection{Calculation of the stiffness}
\label{6v_stiff}

In order to calculate the stiffness for the six-vertex model we map
this model to a loop model, in which the loops are contour loops of
the BCSOS model.

The six-vertex model with vertex weights $a=b=c/2=1$ can be mapped to
a loop model using the break up procedure of Evertz {\em et al.}
\cite{Cevertz}. Every vertex of type $a$ and $b$ is broken up
into two corners of a loop in a unique way, while the $c$ type
vertices are broken up into two possible corners with equal
probability, \Fig{CFbreak}.  Loops generated in this way
are closed and non-intersecting, and they cover all the bonds of the
square lattice. The direction of the arrows around each loop can be
clockwise or counterclockwise, independent of the other loops, giving
a loop fugacity equal to two.

The loops in this model are nothing but {\em contour} loops of the
BCSOS model.  This implies that the correlation function which
measures the probability that two points are on the same loop in the
loop model, is equal to the loop correlation function for a Gaussian 
interface, given by \Eq{loopcor}.
This observation can be used to calculate the value of the stiffness
$K$.

The loop correlation function is equal to the vortex-antivortex
correlation function for vortices of Burgers charge 
$b_{\rm loop} = 2$, as seen
in \Fig{CFvortex}. As shown in the figure, the vortex-antivortex
configuration in the BCSOS model is generated by simply reversing the
arrows along the bonds on one half of the loop. This ``loop trick''
was previously used by Saleur and Duplantier to calculate the
fractal dimension of a percolation hull \cite{Csaleurperc}, and the
fractal dimension of an Ising cluster \cite{CsaleurIsing} in two
dimensions.  Now if we use the formula for the scaling dimension
associated with a vortex-antivortex correlation function, \Eq{dimv},
and we identify it with the universal value of the loop correlation
function exponent $x_{\rm loop}=1/2$, we find
\be{FKexact} K = \frac{\pi}{2} \; .  \ee This value of the stiffness
implies $x({\rm G}_V)=2$ (see \Eq{xV}) which means that the locking
potential $V(h)$ in \Eq{6vfree} is {\em marginal}, and the BCSOS model
is exactly {\em at} the roughening transition. This is in agreement
with the exact solution of the six-vertex model \cite{Cliebwu}, which
predicts a Kosterlitz-Thouless type transition at $a=b=c/2=1$.

It's interesting to note that the exponent $x_{\rm loop}$ has been
measured indirectly by Evertz {\em et al.} \cite{Cevertz}.  Namely, in
their numerical simulations of the six-vertex model, among other
measurements, they measure the average length $<s>$ of a loop of
constant arrow direction as a function of the system size $L$. For
$a=b=c/2=1$ they find \be{FsvsL} <s> \: \sim L^{1.06} \; .  \ee We can
calculate the average length of a loop inside a circle of 
radius $L$, using the loop correlation
function ${\cal G}({\bf x})$; \be{Flcor} <s> \: = \int_0^L \! d^2 \!\:
{\bf x} \; {\cal G}({\bf x}) \; .  \ee If we substitute into this
equation the expression for the loop correlation function,
\Eq{loopcor}, and take $x_{\rm loop}=1/2$,
we find \be{FsvsL2} <s> \: \sim L^{2-2x_{\rm loop}} = L^1 \; , \ee
which is in good agreement with the numerical result,
\Eq{FsvsL}, once systematic errors are taken into account.\footnote
{Ref.~\cite{Cevertz} gave an exponent of 1.060(2), 
but this error includes only statistical errors \cite{Cevertzprivate}.}

\subsection{Symmetry algebra of the six-vertex model}

The effective field theory of the six-vertex model is a Gaussian field 
theory defined by the Euclidean action in \Eq{6vfree2}, and 
compactified on the circle $\Rs/{\cal R}$. This is a conformal field  
theory (CFT) with a stress energy tensor whose holomorphic and  
antiholomorphic components are given by \Eq{stresstensor} where the  
height field has one component.  
The conformal charge is $c=1$, independent of the stiffness $K$. The
modes of the stress-energy tensor are generators of conformal
transformations in the plane, and they form the Virasoro algebra.

For the special value of the stiffness $K=\pi/2$,
that we find for the six-vertex model with vertex weights $a=b=c/2=1$,
the symmetry algebra of the CFT is the chiral $su(2)_{k=1}$ Kac--Moody
algebra.  The currents of the holomorphic half of this algebra are
defined in terms of the holomorphic component of the height $h(w)$,
\Eq{B:holo}, as: \be{Fcurr} J^3 = i \partial h(w) \; , \; \; J^{\pm} =
\: :e^{\pm i \alpha h(w)}: c_{\pm\alpha} \; . \ee
The currents that generate the antiholomorphic part are defined in the
exact same fashion, the only difference being that $h(w)$ is replaced
by $\bar{h}(\bar{w})$.  The reciprocal lattice vector $\alpha=2\pi$ is
chosen so that the currents $J^{\pm}(z)$ have the required conformal
dimension $(1,0)$; see \Eq{confdim}.  The integral multiples of
$\alpha$ generate the {\em root lattice} of an $su(2)$ Lie algebra,
while the reciprocal lattice ${\cal R}^\ast$ can be identified with
the {\em weight lattice} of this algebra; note that the quotient of
the latter with the former is $\Zs_2$, as required for $su(2)$.

The stress energy tensor is given by \Eq{stresstensor}, where the
height field has one component. The stress energy tensor is in the
Sugawara form \cite{Cfuchs}, and therefore the effective field theory
of the six-vertex model with vertex weights $a=b=c/2=1$, is the
$SU(2)_{k=1}$ Wess-Zumino-Witten (WZW) model.  In other words, the
Gaussian field theory that we have proposed as the effective field
theory of the BCSOS model, is the free field representation of this
WZW model.

\section{Three-coloring model}
\label{3colmod}

In this section we show that the symmetry algebra associated with the
three coloring model is the $su(3)_{k=1}$ Kac--Moody algebra. This is
accomplished by mapping the coloring model to an interface model. We
find that the effective field theory of the interface model is
Gaussian with a height field that is compactified on the root lattice
of the $su(3)$ Lie algebra.

\subsection{Definitions}
\label{3col_defs}

The three-coloring model is defined by coloring the bonds of the
honeycomb lattice with three different colors, say $A$, $B$, and $C$,
in such a way that no two bonds of equal color meet at a
vertex.\footnote{ This type of coloring is referred to in the
mathematical literature as an {\em edge coloring}.} Each coloring is
given equal statistical weight, and the partition function can be
written as $Z_0 = \sum_{\cal C} 1$,
where the sum goes over all allowed coloring configurations ${\cal C}$
of the honeycomb lattice.  If we think of the colors as representing
spins in the three-state antiferromagnetic Potts model on the
Kagom\'{e} lattice~\footnote{The bond midpoints of the honeycomb
lattice form the Kagom\'e lattice}, then the coloring model is the
{\em ground state} of the Potts model \cite{Chuse}.

The three coloring model was considered by Baxter \cite{Cbaxter70},
who calculated $Z_0$ exactly.  For a honeycomb lattice of $N$ sites he
found
a non-zero entropy per site $s = \lim_{N\rightarrow\infty} \frac{\ln
Z_0}{N} = 0.1895\ldots \;$.  Below we argue that the three coloring
model is critical, with power law correlations which are described 
in the continuum by the $SU(3)_{k=1}$ Wess-Zumino-Witten model.

\subsection{The interface model}
\label{3col-interface}

The three coloring model can be mapped to a solid-on-solid model which
describes a two dimensional interface in {\em four} spatial dimensions
\cite{Chuse,Cmycomm}. This is accomplished by placing a two component
microscopic height ${\bf z}= (z_1,z_2)$ at the center of each
plaquette of the honeycomb lattice.  The change in ${\bf z}$ when
going from one plaquette to the neighboring one is given by ${\bf A}$,
${\bf B}$, or ${\bf C}$, depending on the color of the bond that is
crossed; see \Fig{C3-idealfig}.  The vectors ${\bf A}$, ${\bf B}$, and
${\bf C}$ point to the vertices of an equilateral triangle: \be{3vecs}
{\bf A} = (-\frac{1}{2},-\frac{\sqrt{3}}{2}) \: , \; \; {\bf B} =
(1,0) \: , \; \; {\bf C} = (-\frac{1}{2},\frac{\sqrt{3}}{2}) \: .  \ee
This ensures that the coloring constraint becomes the zero curl
condition for the height increments, i.e.  the change in height when
going around a plaquette of the triangular lattice, on which the
heights are defined, is $\Delta {\bf z} = {\bf A}+{\bf B}+{\bf C} =
0$.  Each allowed height configuration is given equal statistical
weight.  We will be interested in the long-wavelength fluctuations of
the interface for which we introduce an effective field theory.  We
emphasize that the height fluctuations are entropy driven, in the
sense that they are solely due to the different ways of edge-coloring
the honeycomb lattice with three different colors.

We motivate the long-wavelength theory of the interface model by a
coarse-graining procedure for the microscopic heights ${\bf z}$, which
was described for a general critical ground state in
Sec.~\ref{eff_gen}. Here the general procedure is implemented
as follows: First, we define the ideal states of the three-coloring
model as states in which every elementary plaquette of the honeycomb
lattice is colored with two colors only, Fig.~\ref{C3-idealfig}.
These states are flat, in the sense that they have the smallest
possible variance of the microscopic height. We argue that the free
energy of the coloring model (which is purely entropic in origin) is
dominated by fluctuations around the ideal states.  Namely, the
smallest change on the lattice, that is allowed by the constraints of
the three-coloring model, is an exchange of colors along a {\em loop}
of alternating color (eg. $A$-$B$-$A\ldots$ to $B$-$A$-$B\ldots$). The
ideal states {\em maximize} the number of loops that allow for these
loop exchanges, and it is this property that selects them out. This
entropic selection effect is close in spirit to the ``order by
disorder" effect, introduced by Villain \cite{Cvill}.
 
Second, we divide the honeycomb lattice into domains so that each
domain represents a fluctuation away from a different ideal (flat)
state.  To each domain we assign a coarse grained height ${\bf h}$,
which is equal to the microscopic height averaged over the domain:
       ${\bf h} = \langle {\bf z} \rangle$.
The coarse grained heights associated with different ideal states
form a honeycomb lattice which is the ideal state graph ${\cal I}$, of
the three-coloring model; see \Fig{C3col_graphfig}.  The side of the
elementary hexagon of ${\cal I}$ is $\sqrt{3}/3$, in the units chosen
for the vectors representing the colors; see Eq.~(\ref{3vecs}). Nodes
of ${\cal I}$ that correspond to the {\em same} ideal state form a
triangular lattice with an elementary triangle of side
$\sqrt{3}$. This is the repeat lattice ${\cal R}$ of the
three-coloring model; points on the ideal state graph separated by
vectors in ${\cal R}$ are identified, \be{3ind} {\bf h} \equiv {\bf h}
+ {\cal R} \; .  \ee

Finally, we consider the continuum limit of the interface model in
which the discrete heights, defined over different ideal state
domains, are replaced with a continuously varying height field ${\bf
h}({\bf x}) \equiv (h_1({\bf x}), h_2({\bf x}))$.  Since nodes of the
ideal state graph separated by repeat lattice vectors are
identified~\Eq{3ind}, we take the height field to be {\em
compactified} on the torus,
${\bf h}({\bf x}) \in \Rs^2 / {\cal R}$.
This property of the height field will be important in analyzing the
symmetry algebra of the three coloring model.  The dimensionless free
energy (action) of the interface, which is entropic in origin, is
assumed to be of the form
\begin{equation}
\label{3free1}
f = \int d^{2}{\bf x} \: \left[ \frac{K}{2} (|\nabla h_1|^2 + |\nabla
                         h_2|^2) + V({\bf h}) \right] \; ,
\end{equation}
where $K$ is the stiffness, and $V({\bf h})$ is a periodic potential
with the periodicity given by the ideal state graph,
\begin{equation}
\label{3periodicity}
     V({\bf h} + {\cal I}) = V({\bf h}) \; .
\end{equation}

The free energy $f$ defines the effective field theory of the
three-coloring model; the assumption being made is that it correctly
describes the long-wavelength fluctuations of the microscopic height
${\bf z}$.  The periodic (locking) potential $V({\bf h})$ favors the
heights to take their values on ${\cal I}$, while the first term
represents fluctuations around the flat ideal states.

If the locking potential is irrelevant, in the renormalization group
sense, than the effective field theory of the three-coloring model is
a Gaussian field theory with a dimensionless free energy (action)
given by:
\begin{equation}
\label{3freegauss}
 f = \frac{K}{2} \int d^2 {\bf x} \: (|\nabla h_1|^2 + |\nabla h_2|^2)
\; .
\end{equation}
In the case that the locking potential is relevant, the three-coloring
model will lock into long range order in one of the ideal
states. Which of the two possibilities  is actually 
realized in this model is determined by the value of the stiffness 
$K$ in \Eq{3free1}.

The scaling dimension of the locking potential can be calculated using
the procedure outlined in Sec.~\ref{ops_gen}.  Namely, this operator
has the periodicity of the ideal state graph which, as we saw earlier
in this section, is a honeycomb lattice with a lattice constant
$\sqrt{3}/3$.  The Bravais lattice \cite{CAandM} of the ideal state
graph is a triangular lattice with lattice constant $1$, and its
reciprocal lattice is also triangular, with lattice constant
$\frac{4\pi}{3} \sqrt{3}$. This is also the length of the shortest
vector ${\bf G}_V \in {\cal R}^\ast$ that appears in the Fourier
expansion of $V({\bf h})$. Therefore, from \Eq{dim}, the scaling
dimension of the locking potential is \be{3VdimK} x({\bf G}_V) =
\frac{4\pi}{3 K} \; .  \ee
The locking potential will be irrelevant, in the renormalization group
sense, if $x({\bf G}_V)>2$, or if the stiffness \be{3Kineq} K <
\frac{2\pi}{3} \; .  \ee For $K =\frac{2\pi}{3}$ the three-coloring
model, in the interface representation, undergoes a roughening
transition; for $K>\frac{2\pi}{3}$ the model will lock into long range
order in one of the six ideal states.

\subsection{Calculation of the stiffness}

Here we examine the correlation function that measures the probability
that two points on the honeycomb lattice are on the same loop of {\em
alternating} color.  The crucial observation that these loops are
contour loops for a particular component of the microscopic height,
will allow us to calculate the stiffness {\em exactly}.

If we choose two colors, for instance $A$ and $C$, then due to the
edge coloring constraint there will be a $A$-$C$-$A$-$\ldots$ loop
passing through every vertex of the honeycomb lattice.  The
correlation function that measures the probability that two points,
say one at $0$ and the other at ${\bf R}$, belong to the {\em same}
$AC$ loop is \be{3loopcor} {\cal G}({\bf R}) = \frac{Z({\bf R})}{Z_0}
\; .  \ee The restricted partition function $Z({\bf R})$ is simply the
number of colorings with an $AC$ loop passing through $0$ and ${\bf
R}$, while $Z_0$ is the total number of colorings.  Now, if we
exchange the two colors on the loop along one half of the loop going
from $0$ to ${\bf R}$, then this will generate a vortex and an
antivortex in the interface model at these two points; see
\Fig{C3coldefectsfig}.
If we do this for all the configurations entering $Z({\bf R})$, then
${\cal G}({\bf R})$ becomes the vortex-antivortex correlation
function. The Burgers charges associated with these vortices are:
\be{3Bcharge} {\bf b}_{\rm loop} = \pm ({\bf A} - {\bf C}) = \pm (0,
-\sqrt{3}) \ee and the loop correlation function is given by
\be{3Loopcor} 
{\cal G}({\bf R}) \sim R^{-2 x_v({\bf b}_{\rm loop})} \; .  
\ee 
The exponent $x_v({\bf b}_{\rm loop})$ is the scaling 
dimension of a magnetic type operator, and from Eqs.~(\ref{dimv}) 
and~(\ref{3Bcharge}) we find \be{3x1K} x_v({\bf b}_{\rm loop}) = 
\frac{3 K}{4 \pi} \; .  \ee

In the interface representation of the three-coloring model loops of
alternating color become loops of constant height.  This is most
readily understood from an example: take an $AC$ loop and consider the
points at the centers of the hexagonal plaquettes along the inside of
the loop (Fig.~\ref{C3coldefectsfig}).  These points are separated by
$B$ colored bonds, and therefore the component of the microscopic
height ${\bf z}\cdot{\bf e}_2$ is unchanged as we go around the loop,
since the projection ${\bf B}\cdot{\bf e}_2=0$; $\{{\bf e}_1,{\bf
e}_2\}$ are the orthonormal basis vectors in the height space.  In
general, every pair of colors defines a loop of alternating colors on
the lattice, which is also a contour line of the component of ${\bf
z}$ in the direction which is perpendicular to the vector representing
the third color.

The probability that two points separated by $R$ belong to the same
contour loop of a Gaussian interface scales as $R^{-1}$ for large $R$
(see Section~\ref{contours_gen}, \Eq{loopcor}). Therefore, \be{3colx1}
2 x_v({\bf b}_{\rm loop}) = 1 \; , \ee and from \Eq{3x1K}
\be{3colKnum} K = \frac{2 \pi}{3} \ee is the {\em exact} value of the
stiffness.

If we compare \Eq{3colKnum} with \Eq{3Kineq} we can conclude that the
three-coloring model in the interface representation is exactly {\em
at} the roughening transition.  This observation justifies the
Gaussian form of the free energy, \Eq{3freegauss}, and it also follows
from Baxter's exact solution of the three coloring model
\cite{Cbaxter70}, as was shown by Huse and Rutenberg \cite{Chuse}.

\subsection{The symmetry algebra of the three-coloring model}

In this section we analyze the conformal field theory of the three
coloring model and we show that it is given by an $SU(3)_{k=1}$
Wess--Zumino--Witten model \cite{Cwitt}, as conjectured by Read
\cite{Cread}.

The effective field theory of the three-coloring model is a Gaussian
field theory, \Eq{3freegauss}, compactified on the torus $\Rs^2/{\cal
R}$, \Eq{3ind}.  This is a conformal field theory with conformal
charge $c=2$, independent of the value of the stiffness $K$.

The symmetry algebra of the effective field theory of the
three--coloring model is the chiral $su(3)_{k=1}$ Kac-Moody algebra.
In order to show this we explicitly construct the currents of the
holomorphic half of the algebra from the holomorphic component of the
height field ${\bf h}(w)$, \Eq{B:holo}, using the Frenkel--Kac--Segal
vertex operator construction of level one Kac--Moody algebras
\cite{Cfks}.  The construction of the antiholomorphic half of the
chiral algebra follows in the same fashion, the only difference being
that ${\bf h}(w)$ is replaced by the antiholomorphic component of the
height field $\bar{{\bf h}}(\bar{w})$.

There are eight conserved currents, equal in number to the dimension
of the $su(3)$ Lie algebra. The two currents that correspond to the
Cartan subalgebra of $su(3)$ are:
\be{3Cartan} H_1(w) = i \partial h_1(w) \; , \; \; \; H_2(w) = i
   \partial h_2(w) \; .  \ee The remaining six currents, which are the
   raising and lowering operators associated with the positive roots
   $\ba_j$, are the vertex operators \be{3CarWey} J_{\pm{\ba}_j}(w) =
   \: :e^{\pm i \ba_j \cdot {\bf h}(w)}: c_{\pm \ba_j} \; \; (j=1,2,3)
   \; .  \ee

The positive roots are vectors in the reciprocal lattice ${\cal
R}^\ast$ shown in \Fig{C3rootsfig}; they are of length $|{\ba}_j| =
\frac{4\pi}{3}\sqrt{3}$, in units chosen for ${\bf A}$, ${\bf B}$, and
${\bf C}$ in \Eq{3vecs}.  From \Eq{confdim} their conformal dimension
is $(1,0)$ as expected for a current field.
The vectors $\ba_j$ generate a triangular lattice which is the {\em
root} lattice of an $su(3)$ Lie algebra.  The quotient of the
reciprocal lattice, which we identify with the {\em weight} lattice of
$su(3)$, and the root lattice, is $\Zs_3$, as required of an $su(3)$
Lie algebra.

Finally, since the stress energy tensor, \Eq{stresstensor}, is in the
Sugawara form \cite{Cfuchs}, we can conclude that the effective filed
theory of the three-coloring model is the $SU(3)_{k=1}$
Wess-Zumino-Witten (WZW) model.

The existence of a hidden $SU(3)$ symmetry in the three-coloring model
was previously shown by Read \cite{Cread}, who mapped the
three-coloring model to a lattice model in which the symmetry is
explicit. Our approach is very different, it shows the emergence of an
$su(3)_{k=1}$ Kac--Moody algebra in the continuum, and it is based on
the interface representation of the three-coloring model.

\section{The four coloring model}
\label{4colmod}

In this section we study the ground state of of the four-state
antiferromagnetic Potts {\em vertex} model on the square lattice. This
model was first introduced by Read~\cite{Cread} as a generalization of
the three-state antiferromagnetic Potts model on the Kagom\'{e}
lattice.  We have recently completed a detailed study of the ground
state of this Potts model \cite{CKHPRB}, both analytically and using
Monte-Carlo simulations, and evidence was found in support of the
claim that the ground state is indeed critical.  Here we focus on the
symmetry of the field theory that describes the continuum limit of this
critical ground state,
and we find that it is given by the chiral $su(4)_{k=1}$ Kac--Moody
algebra.

\subsection{Definitions}
\label{4defs}

In this section we introduce the four-coloring model as the ground
state of the antiferromagnetic four-state Potts vertex model on the
square lattice.

Our starting point is the antiferromagnetic Potts vertex model given
by the Hamiltonian
\begin{equation}
\label{hamiltonian}
 {\cal H} = |J| \sum_{\bf x} \sum_{\stackrel{\scriptstyle i,j =
1}{i<j}}^{4} \delta (\sigma_i({\bf x}), \sigma_j({\bf x}) )
\end{equation}
where the Potts spins $\sigma_i({\bf x}) (i=1,2,3,4)$, live on 
the four 
bonds of the square lattice which share the same vertex 
${\bf x}$;
each spin can be in one of four possible states labeled $A$, $B$,
$C$, and $D$. This Hamiltonian associates an energy penalty $|J|$
to having 
equal spins on two vertex-sharing bonds 
of the square lattice.
An alternative representation of this model, given by
Read~\cite{Cread}, involves the ``crossed-square" lattice in which
diagonal bonds are drawn on every other square plaquette, so that the
crossed plaquettes form a checkerboard pattern. In this
representation the Potts spins live on the {\em vertices} of the
crossed-square lattice and have nearest neighbor antiferromagnetic
interactions.

At zero temperature, the only allowed spin configurations are ones for
which $\{ \sigma_i({\bf x}): \: i=1,2,3,4 \}$ $=$ $\{ A, B,$ $ C, D \}
$ for every ${\bf x}$. With $\{ \sigma_i({\bf x}): \: i=1,2,3,4 \}$ we
denote the {\em set} of Potts spins on the four bonds at the vertex
${\bf x}$, while by the {\em ordered set} $(\sigma_1({\bf x}),
\sigma_2({\bf x}),$ $\sigma_3({\bf x}),\sigma_4({\bf x}))$ we 
will denote
the particular arrangement of spins at ${\bf x}$.

The ground state ensemble has an extensive entropy.  Namely, the
ground state entropy per site, defined as~\cite{Cliebwu} 
$s = \lim_{N \rightarrow \infty} \frac{1}{N} \ln(Z_0)$,
where $Z_0$ is the number of ground states and $N$ the number of
sites, is non-zero.  This can be easily verified by examining the
state given in Fig.~\ref{Cideal4}: In every $A$$B$-plaquette the spins
$A$ and $B$ can be exchanged independently of the other
plaquettes. This gives rise to $2^{N/4}$ states, which puts a lower
bound on the entropy per site at $s >\ln2/4$.

If we think of the four Potts spins as four colors then the ground
states correspond to the four-coloring model of the square lattice.
In the four-coloring model each bond of the square lattice is colored
with one of four different colors $A$,$B$,$C$ or $D$, with the
constraint that at each vertex all four colors meet.  All such
configurations are given the same statistical weight.

\subsection{The interface model}
\label{4col-interface}

Here we describe the mapping of the four-coloring model to an
interface model, and we propose an effective field theory that
describes the long-wavelength fluctuations of this interface.  Further
details of this mapping, and an analysis of the four-coloring
operators in the interface representation, can be found in
\cite{CKHPRB}.

We define a height mapping by placing a {\em three-component}
microscopic height ${\bf z} \in \Zs^3$ at the center of each
elementary square (plaquette). The change in ${\bf z}$, when going
from one plaquette to a neighboring one, is given by $\Delta{\bf
z}=\bs({\bf x})$, where $\bs({\bf x})$ is the color of the bond that
is crossed;
see Fig.~\ref{Cideal4}.  The four possible color values that $\bs({\bf
x})$ can take are represented by vectors pointing to the vertices of a
tetrahedron:
\begin{eqnarray}
\label{vectors}
    {\bf A} = (-1,+1,+1), \; \; \; {\bf B} = (+1,+1,-1), \nonumber \\
    {\bf C} = (-1,-1,-1), \; \; \; {\bf D} = (+1,-1,+1) \; .
\end{eqnarray} 
For a given configuration of colors, the set of microscopic heights
defines a two dimensional interface in five dimensions. Each allowed
microscopic height configuration is given equal statistical weight.

In order to define an effective field theory for the above described
interface model, we introduce a coarse-graining procedure for the
microscopic heights, following the general procedure outlined in
Section~\ref{eff_gen}.  First, we define ideal states of the
four-coloring model as states in which every plaquette is colored by
two colors only; see Fig.~\ref{Cideal4}.  There are $24=4!$ ideal
states related to each other by lattice symmetries, and each
corresponds to a different permutation of the 4 colors.  These states
are flat, in the sense that the variance of the microscopic height is
minimum. Furthermore, they have the important property that they are
entropically selected, in the sense that ideal states allow for the
maximum number of color rearrangements consistent with the ground
state constraints.
This point is crucial. Namely, if we wish to change the color of a
bond, then the smallest change we can perform on the lattice, without
violating the constraints of the four-coloring model, is an exchange
of two colors along a loop of alternating color. Just as in the
three-coloring model, the ideal states {\em maximize} the number of
loops of alternating color.

Second, we replace the original model with a coarse grained version
where the lattice is split into domains, such that in each domain
fluctuations occur about a different ideal state.  With each
ideal-state domain we associate a coarse grained height ${\bf h}$,
which is given by the average microscopic height in that domain, ${\bf
h} = <{\bf z}>$.

Third, we define the ideal state graph ${\cal I} \subset
{\Rs}^3$. Every node of ${\cal I}$ represents an ideal state, and its
position in ${\Rs}^3$ is given by the coarse grained height ${\bf h}$
of the ideal state it represents.  The 24 {\em different}
ideal states form a truncated octahedron with a side of length
$\sqrt{2}/2$ in the units chosen for the vectors ${\bf A}$, ${\bf B}$,
${\bf C}$ and ${\bf D}$; see Fig.~\ref{Coctahedron}. These truncated
octahedra are arranged {\em periodically} in a face centered cubic
(FCC) lattice, which is the repeat lattice $\cal R$, to form the full
ideal state graph. The side of the conventional cubic cell of ${\cal
R}$ is $4$.  Since ${\bf h}$ and ${\bf h} + \cal R$ represent the same
ideal state on the ideal state graph, we can identify them
\begin{equation}
\label{compactify}
{\bf h} \equiv \: {\bf h} + {\cal R} \; .
\end{equation} 
In other words the height is compactified on the three-torus
$\Rs^3/{\cal R}$.

Finally, we consider the long-wavelength limit of the interface model,
where the heights defined over particular ideal state domains are
replaced with a continuously varying height field ${\bf h}({\bf x})
\equiv (h_1({\bf x}), h_2({\bf x}), h_3({\bf x}))$.  The dimensionless
free energy (action) of the interface, which is entropic in origin, is
assumed to be of the form:
\begin{equation}
\label{4free}
f = \int d^2 {\bf x} \: \left[\frac{K}{2} (|\nabla h_1|^2 + |\nabla
          h_2|^2 + |\nabla h_3|^2) + V({\bf h}) \right],
\end{equation}
where $V({\bf h})$ is a periodic potential with the periodicity given
by the ideal state graph,
\begin{equation}
\label{periodicity}
     V({\bf h} + {\cal I}) = V({\bf h}) \; .
\end{equation}

The free energy $f$ defines an effective field theory of the
four-coloring model.  The periodic potential $V({\bf h})$, which is
usually referred to as the {\em locking potential}~\cite{CJKKN},
favors the heights to take their values on ${\cal I}$, while the first
term represents fluctuations around the flat ideal states.

The locking potential is periodic in height space with the periodicity
of ${\cal I}$, and therefore its scaling dimension is given by
\Eq{dim}, i.e.,
\begin{equation}
\label{locking}
 x({\bf G}_V) = \frac{\pi}{2 K} \; ;
\end{equation}
$|{\bf G}_V| = \sqrt{2} \: \pi$ is the shortest reciprocal lattice
vector appearing in the Fourier expansion of $V({\bf h})$.  The
magnitude of ${\bf G}_V$ can be deduced from Fig.~\ref{Coctahedron},
which shows that the Bravais lattice of the ideal state graph is a
body centered cubic (BCC) lattice, with a conventional cubic cell
whose side is of length $2$. The reciprocal of this lattice is an FCC
lattice with a conventional cubic cell of side $2 \pi$, and
consequently the magnitude of its shortest lattice vector (${\bf
G}_V$) is $2 \pi/\sqrt{2}$.

If the stiffness $K$ satisfies $K<\pi/4 $, then the locking potential
in Eq.~(\ref{4free}) is irrelevant in the renormalization group sense,
i.e., $x({\bf G}_V)>2$, and the four-coloring model is described by a
Gaussian free energy (action)
\begin{equation}
\label{4freegauss}
 f = \frac{K}{2} \int d^2 {\bf x} \: (|\nabla h_1|^2 + |\nabla h_2|^2
          + |\nabla h_3|^2) \; .
\end{equation}
In the case that the locking term is relevant ($K>\pi/4$, i.e., $x({\bf
G}_V)<2$), the four-coloring model will lock into long range order in
one of the ideal states.  We will see later that, just as in the 
six-vertex model with $a=b=c/2=1$, and the three-coloring model,
the four coloring model is {\em at} the roughening transition.

\subsection{Calculation of the stiffness}

In perfect analogy with the six-vertex model and the three-coloring
model, we calculate the stiffness of the four-coloring model using
the loop trick. Namely, we express the loop correlation function
exponent $x_{\rm loop}$, for contour loops on a Gaussian interface, in
terms of the stiffness, and then we use the universal value of this
exponent, $x_{\rm loop}=1/2$ (\Eq{loopcor}).

We start by calculating the loop correlation function ${\cal G}({\bf
R})$, for loops of alternating color in the four-coloring model, in
the familiar way~\cite{Cnienrev}.  If we denote the number of
configurations with a loop of alternating color passing through points
$0$ and ${\bf R}$ as $Z({\bf R})$, then ${\cal G}({\bf R})$ can be
written as
\begin{equation}
\label{loopcordef}
{\cal G}({\bf R}) = \frac{Z({\bf R})}{Z_0} \; ,
\end{equation}
where $Z_0$ is the total number of configurations; $Z_0$ is also the
partition function, since all configurations in the four-coloring
model have equal statistical weight.  Now, if we exchange the two
colors on the loop along one half of the loop going from $0$ to ${\bf
R}$, then this generates a vortex and an antivortex in the interface
model, at these two points.  For example, if the loop consists of
alternating $A$ and $B$ colored bonds, then the color configuration,
after flipping half the loop, is $(C,D,A,A)$ at one end, and
$(B,B,C,D)$ at the other (see Fig.~\ref{Cdefects}).  Using the height
rule we find that the Burgers vectors associated with these
``defects'' are
\begin{equation}
\label{burgers1}
\pm {\bf b}_{\rm loop} = \pm ({\bf A} - {\bf B}) = \pm (-2,0,2) \; .
\end{equation}
Therefore, the loop correlation function is equal to the probability
of having a vortex-antivortex pair of Burgers charge $\pm {\bf b}_{\rm
loop}$ separated by ${\bf R}$, in the interface model.  The
vortex-antivortex correlation function scales with distance
as~\cite{CJKKN}
\begin{equation}
\label{c6:loopcorr}
{\cal G}({\bf R}) \sim |{\bf R}|^{-2 x_v({\bf b}_{\rm loop})} \; ,
\end{equation}
where the exponent $x_v({\bf b}_{\rm loop})$ is given by \Eq{dimv}.

In the interface representation, loops of alternating color become
loops of constant height.  This is most readily understood from an
example: take an $AB$ loop, and consider the points at the centers of
the plaquettes along the inside of the loop (Fig.~\ref{Cdefects}).
These points are separated by $C$ and $D$ bonds only, and therefore
the component of the microscopic height ${\bf z}$ in the ${\bf e}_1 -
{\bf e}_3$ direction is unchanged as we go around the loop. This
follows from the fact that the projections of both ${\bf C}$ and ${\bf
D}$ are zero in this direction; $\{{\bf e}_1,{\bf e}_2,{\bf e}_3\}$
are the orthonormal basis vectors in the height space.  In general,
every pair of colors defines a loop of alternating color, 
which is a contour line of the component of ${\bf z}$ in the direction
which is perpendicular to the vectors representing the other two
colors.

The observation that loops of alternating color are contour lines
leads to the equation \be{xlsamex1} x_v({\bf b}_{\rm loop}) = x_{\rm
loop} = \frac{1}{2} \; .  \ee By making use of \Eq{dimv} and
\Eq{burgers1} in the above equation, we find for the exact value of
the stiffness
\begin{equation}
\label{Kexact}
 K = \frac{\pi}{4} \; .
\end{equation}
As advertised earlier, the exact value of $K$ is such that the locking
potential is marginal,
and we conclude that the interface model is {\em at} the roughening
transition. The exact scaling dimensions of operators in the
four-coloring model are completely determined by $K$. It is worth
noting that we have measured $K$ in Monte-Carlo simulations of the
four-coloring model \cite{CKHPRB} and we find $K^{-1}=1.28\pm 0.01$,
in excellent agreement with the exact result
$K^{-1}=1.273\ldots$, \Eq{Kexact}.

\subsection{The symmetry algebra of the four-coloring model}

We have seen that the four-coloring model can be mapped onto a
Gaussian interface, Eq.~(\ref{4freegauss}), with a three-component
height that is {\em compactified} on the repeat lattice ${\cal R}$,
Eq.~(\ref{compactify}).  Thus, the conformal field theory (CFT) that
emerges in the scaling limit is a rather simple one. It corresponds to
three massless free bosons ${\bf h} = (h_1,h_2,h_3)$ compactified on
the face-centered cubic (FCC) lattice ${\cal R}$.  The conformal
charge of this CFT is $c=3$.

The root lattice of an $su(4)$ Lie algebra is also an FCC
lattice~\cite{Cfuchs}. Therefore, if the radius of compactification of the
height field has the correct value, then the CFT has an infinite symmetry
larger than the usual conformal symmetry, and it is given by the
$su(4)_{k=1}$ Kac-Moody algebra.  The radius of
compactification is equivalent to the stiffness of the
interface, which was calculated exactly in the previous section; see 
\Eq{Kexact}.

The Kac-Moody algebra associated with the scaling limit of the
four-coloring model is chiral.
The currents of the holomorphic half of the chiral algebra, that
correspond to the Cartan subalgebra of $su(4)$ are \be{4Cartan} H_1(w)
= i \partial h_1(w) \; , \; \; H_2(w) = i \partial h_2(w) \; , \; \;
H_3(w) = i \partial h_3(w) \; , \ee while the remaining twelve,
associated with the raising and lowering operators, are given by
\be{4CarWey} J_{{\ba}_j}(w) = \: :e^{i \ba_j \cdot {\bf h}(w)}:
c_{\ba_j} \; \; (j=1,2,\ldots,12) \; .  \ee
The currents of the antiholomorphic half of the chiral algebra are
also given by Eqs.~(\ref{4Cartan}) and~(\ref{4CarWey}), only now ${\bf
h}(w)$ is replaced with the antiholomorphic component
 of the height field
$\bar{{\bf h}}(\bar{w})$.  The twelve {\em roots} ${\ba}_j$  are
$(1,1,0)$-type vectors in the reciprocal lattice, which we identify
with the {\em weight} lattice of the $su(4)$ Lie algebra; 
see \Fig{4roots}.  The lattice
generated by the roots is the root lattice and the quotient of the
weight lattice by the root lattice is $\Zs_4$, as required of an
$su(4)$ Lie algebra.  Here we see the special role played by the
stiffness K. Its value, \Eq{Kexact}, ensures that the conformal
dimension of the fields $J_{{\ba}_j}(w)$ is $(1,0)$, \Eq{confdim},
which is necessary if $J_{{\ba}_j}(w)$ are to be conserved currents.

The idea that the scaling limit of the four-coloring model is given by
the $SU(4)_{k=1}$ Wess-Zumino-Witten (WZW) model was first put forward
by N. Read~\cite{Cread}.  He showed that the four-coloring model is
equivalent to a lattice model with explicit $SU(4)$ symmetry and went
on to conjecture that the scaling limit is given by a WZW model.  From
the interface representation of the four-coloring model, we have shown
the emergence of a chiral $su(4)_{k=1}$ Kac--Moody algebra in the
scaling limit. Taking into account the fact that the stress-energy
tensor is in the Sugawara form (see \Eq{stresstensor}) proves the
conjecture put forward by Read.

\section{Summary and remarks}

In this paper we have developed a fairly simple geometric approach to
analyzing the symmetry properties of discrete spin models with
critical ground states. This approach combines certain ideas developed
in the study of interface models and the roughening transition, with
the vertex operator construction of level-one Kac--Moody algebras.

The first step is to map a critical ground state to an interface 
model. The stiffness of the interface plays a central role as it
determines the values of the scaling dimensions of all the operators
in the critical ground state. In order to calculate the stiffness we
turn to the loop correlation function ${\cal G}({\bf x})$, which is
defined as the probability that two points, separated by ${\bf x}$,
belong to the same contour loop of a Gaussian (rough) interface. 
In the critical ground states studied here, we were able to identify 
contour loops of the associated interface models,
and express ${\cal G}({\bf x})$ as a vortex-antivortex
correlation function.
{}From the universal, i.e., stiffness {\em independent} 
value of the loop correlation
function exponent $x_{\rm loop}=1/2$, the stiffness of the interface
could be calculated exactly. This was accomplished by identifying $x_{\rm
loop}$ with the vortex-antivortex correlation function exponent, which is
stiffness {\em dependent}.  The value of the stiffness, and the fact that
the heights are compactified on a lattice, conspire to give a
Kac--Moody algebra as the symmetry algebra of the conformal field
theory, which describes the long-wavelength fluctuations of the
interface. In particular, for the six-vertex model with
$a=b=c/2=1$, the three-coloring model, and the four-coloring 
model, we found the value of the stiffness to be such, that  
the associated interface models, in all three cases, are exactly 
at the roughening transition. Furthermore, we were able to show 
that the effective field theories of these interface models 
are the $SU(2)_{k=1}$, $SU(3)_{k=1}$, and $SU(4)_{k=1}$
Wess-Zumino-Witten model, respectively. 
     
A couple of final remarks are in order. The three models we have 
studied here share a common property, namely, each one 
can be mapped 
to a fully packed loop (FPL) model. The building blocks
of an FPL model are loops that run along the bonds of a
two-dimensional lattice, with the constraints that every vertex of the
lattice belongs to a loop, and the loops do not intersect.  The
six-vertex model maps to an FPL model on the square lattice via the
break up procedure described in Section~\ref{6v_stiff}. Here the loops
are loops of constant arrow direction.
The three-coloring model maps to the fully packed loop model on the
honeycomb lattice \cite{Cbloteprl} where the loops are loops of
alternating color, say $A$ and $B$ \cite{Cmycomm}.  Finally, the
four-coloring model is equivalent to a fully packed loop model with
{\em two} loop flavors, say $AB$ and $CD$ \cite{CKHPRB}; in this case
we allow for two loops of different flavor to intersect. 

The fugacity of loops in all of the above mentioned models is two,
corresponding to the two possible states each loop can be in; there
are two choices for the arrow direction along a loop in the 
six-vertex model, and two ways of coloring a loop of alternating color,
in the coloring models.  Lowering
the loop fugacity leaves the loop models critical; on the level of the
effective field theory of the loop model, this corresponds to
perturbing the appropriate Wess-Zumino-Witten model
by an exactly marginal operator and introducing a background charge 
\cite{CKondevDeGier}. 
Increasing the loop fugacity above two, on the 
other hand, leads to a finite correlation
length which is roughly the size of the largest loop.  
In the limit of vanishing fugacities these loop models define different
variants of the self-avoiding walk problem. 

Recently Batchelor {\em et al.} \cite{Cbatch} found  a Bethe 
ansatz solution of the fully packed loop model on the honeycomb
lattice. They calculated the so-called watermelon dimensions,
and the conformal charge, as a function of the loop fugacity. 
We have been able to reproduce these 
results from an analysis based on the mapping to an interface model 
\cite{CKondevDeGier}, and were also able to calculate the so-called 
temperature dimension which was found numerically by Bl\"ote and 
Nienhuis \cite{Cblote}, 
and which does not appear in the Bethe ansatz solution. 
Analogous calculations can be done for the other loop models, and the
conformal charge, as well as
the complete spectrum of scaling dimensions along
the critical line, can be determined. 

It is possible to define $N$-coloring models as a generalization
of the 3- and 4-coloring model. For $N = 6$,
the simplest realization is given by the 6-coloring model on the triangular
lattice, where the bonds are colored with six different colors so
that at each vertex all six are represented. It would be interesting
to see if these models have an $su(N)$ symmetry associated with
them. We hope to address this problem in the near future, both numerically
and analytically.

We  would like to acknowledge very useful  
discussions with A. LeClair, C. Zeng, 
A.A. Ludwig, and N. Read. This work was supported by the NSF through 
Grant No. DMR-9214943.

\begin{figure}
\caption
{
\label{Csketch}
The construction of a fluctuating interface equivalent to the ground
state ensemble of a discrete spin model: a) The microscopic spin
configuration is broken up into ideal-state domains. b) Each ideal
state domain is assigned a coarse grained height equal to the average
microscopic height of the domain. c) Finally, the discrete heights are
replaced by a continuous height field ${\bf h}({\bf x})$; the
interface is assumed to be rough. }
\end{figure}

\begin{figure}
\caption{
\label{C6vweights}
Vertex configurations of the six-vertex model, and their respective
Boltzmann weights, $a$, $b$, and $c$.}
\end{figure}
\begin{figure}
\caption{
\label{C6v_ideal}
One of the two symmetry related ideal states of the six-vertex
model. The heights are defined at the vertices of the dual lattice via
the height rule. Note that the ideal state is macroscopically flat.}
\end{figure}
\begin{figure}
\caption{
\label{C6v_ideal_graph}
The ideal state graph of the six-vertex model. The ideal states are
represented by vertex configurations at the origin, 
and the coarse grained height
$h$ assigned to each one is equal to the average microscopic height.  The
graph is a one-dimensional lattice with a lattice spacing equal to
$1$.  Every two vertices the ideal states repeat, and therefore the
repeat lattice vectors are integer multiples of $2$.}
\end{figure}

\begin{figure}
\caption
{\label{CFbreak}
The six-vertex model with vertex weights $a=b=c/2=1$ can be turned
into a model of non-intersecting loops that cover all the bonds of the
square lattice. Every vertex configuration is broken up into a loop
corner: for the a- and b-type vertices this is done in a unique way,
while the two break up possibilities for the c-type vertex are assumed
equally likely.  This break up procedure gives every one of the eight
possible loop configurations at a vertex equal statistical weight.
The loops generated in this way are {\em contour} loops for the BCSOS
model.}
\end{figure}
\begin{figure}
\caption
{\label{CFvortex}
Vortices in the BCSOS model correspond to violations of the ice rule
in the six-vertex model. A vortex-antivortex pair (circled) appears
when arrows along one half of a loop of constant arrow direction, are
flipped. These loops are contour lines in the BCSOS model. The Burgers
charge of the vortices is $b_1=\pm 3$.}
\end{figure}
\begin{figure}
\caption
{
\label{C3-idealfig}
One of six symmetry related ideal states of the three-coloring
model. In an ideal state all the plaquettes are colored with two
colors only. The microscopic heights ${\bf z}$ are defined at the
centers of the plaquettes, and the change in ${\bf z}$, when going
from one plaquette to the neighboring one, is determined by the color
of the crossed bond. The ideal state is macroscopically flat, in the
sense that the variance of the microscopic height is minimal.  }
\end{figure}
\begin{figure}
\caption
{\label{C3col_graphfig}
The ideal state graph of the three coloring model is a honeycomb
lattice: each vertex is associated with a particular ideal state, and
the six different ideal states form a hexagonal plaquette. The ideal
states are labeled by the color configuration
$(\sigma_1,\sigma_2,\sigma_3)$ of the bonds around the origin.  The
vertices in the ideal state graph that correspond to the {\em same}
ideal state (say $(C,B,A)$) form a triangular lattice which is the
repeat lattice of the three-coloring model.  }
\end{figure}
\begin{figure}
\caption
{\label{C3coldefectsfig}
Elementary defects in the three-coloring model are associated with
loops of alternating color: exchanging the two colors (A and C) along
one half of the loop (shown in bold) will generate defects (circled),
i.e., violations of the edge coloring constraint, at the two ends. In
the interface representation these defects become vortex-antivortex
configurations of the height. The loop correlation function is equal
to the vortex-antivortex correlation function.
}
\end{figure}
\begin{figure}
\caption
{
\label{C3rootsfig}
The three positive roots are identified with the second shortest
reciprocal lattice vectors; the reciprocal lattice is a triangular
lattice, with the lattice spacing $\frac{4\pi}{3}$.  The vertex
operators associated with the roots
are currents of the $su(3)_{k=1}$ Kac--Moody algebra.  }
\end{figure}
\begin{figure}
\caption
{
\label{Cideal4}
One of 24 symmetry related ideal states of the four-coloring
model. Using the height rule defined in the text, every plaquette is
assigned a microscopic height ${\bf z}$. Note that in the ideal state
the height describes, on average, a flat interface with a fast
modulation of the microscopic height.  }
\end{figure}
\begin{figure}
\caption{
\label{Coctahedron}
The ideal state graph of the four-coloring model.  The vertices that
correspond to the 24 different ideal states form a truncated
octahedron.  The full ideal state graph corresponds to a periodic
tiling of space with these octahedra. They are arranged in a
face-centered cubic lattice, which is the repeat lattice of the
four-coloring model.}
\end{figure}
\begin{figure}
\caption
{\label{Cdefects}
A vortex-antivortex pair (circled) in the four-coloring model; the
pair belongs to a loop of alternating color (bold).  The loop
correlation function is equal to the vortex-antivortex correlation
function. This can be seen by exchanging the colors ($A$ and $B$)
along one part of the loop between the circled vertices.}
\end{figure}
\begin{figure}
\caption
{\label{4roots}
The twelve root vectors are $(1,1,0)$-type
lattice vectors in the BCC lattice ${\cal R}^\ast$;
they lie in planes perpendicular to $(1,0,0)$-type vectors.
The lattice ${\cal R}^\ast$ is identified with the
weight lattice of the $su(4)$ Lie algebra, while the 
root vectors generate the root lattice. 
Vertex operators
associated with the root vectors are currents of the $su(4)_{k=1}$
Kac--Moody algebra. 
}
\end{figure}

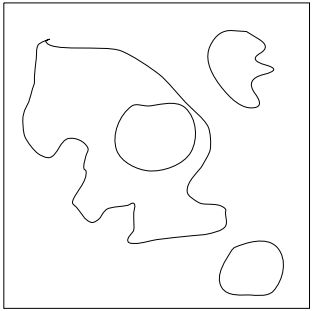
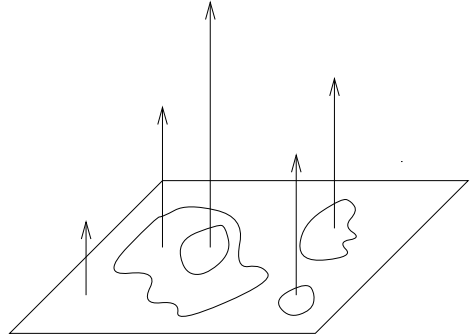
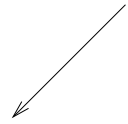
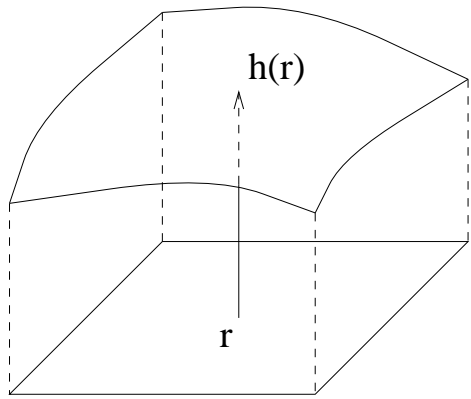
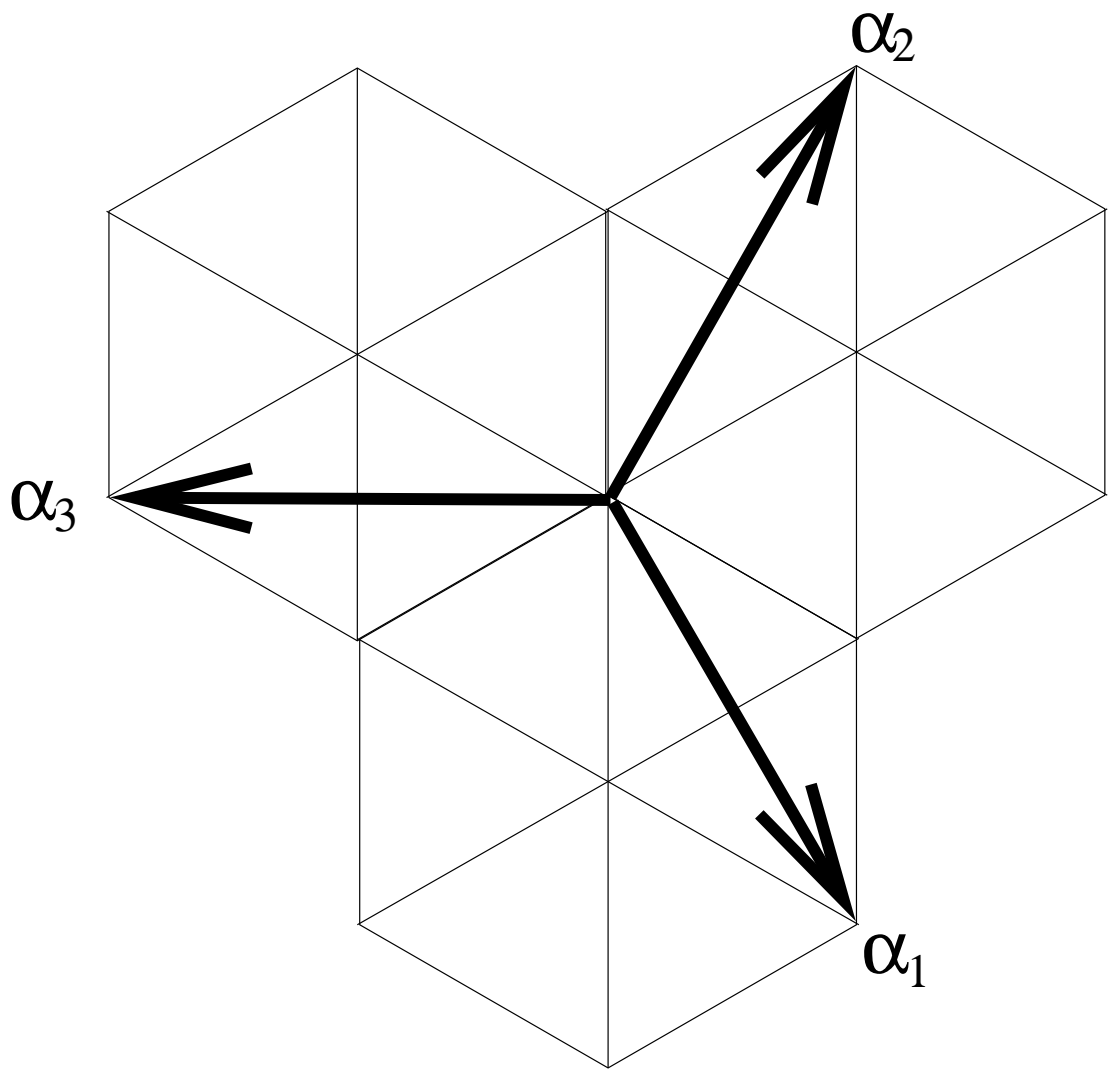
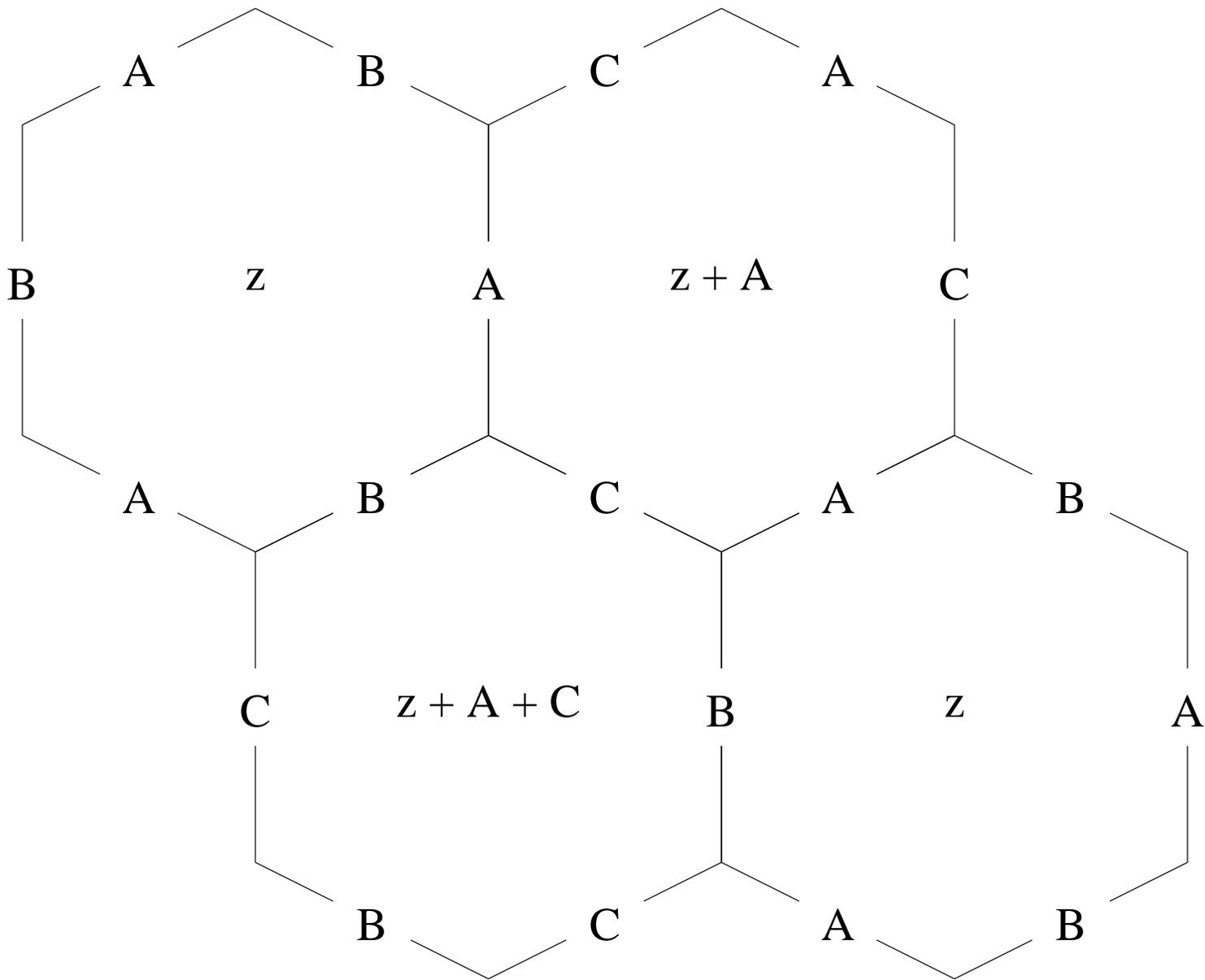
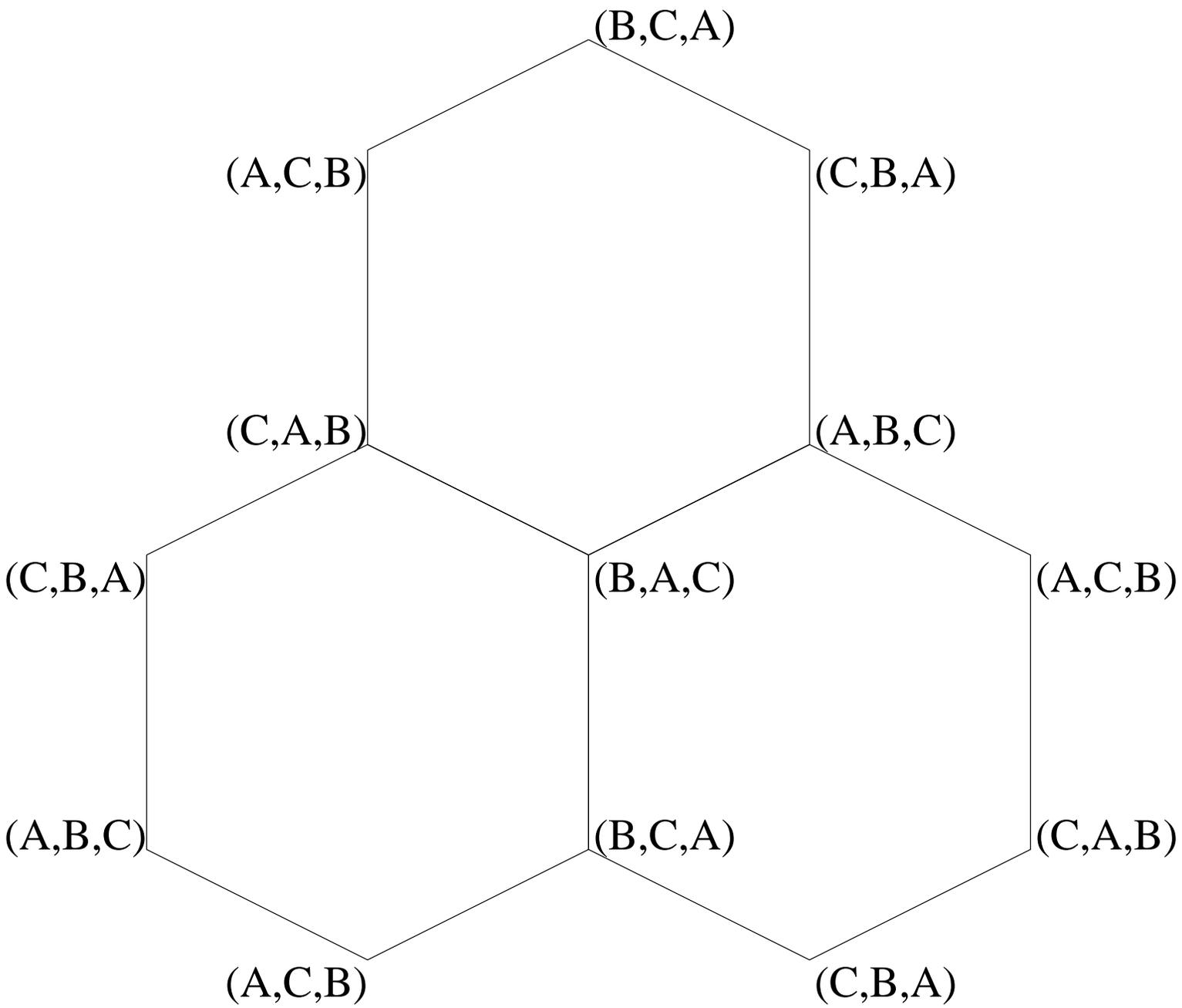
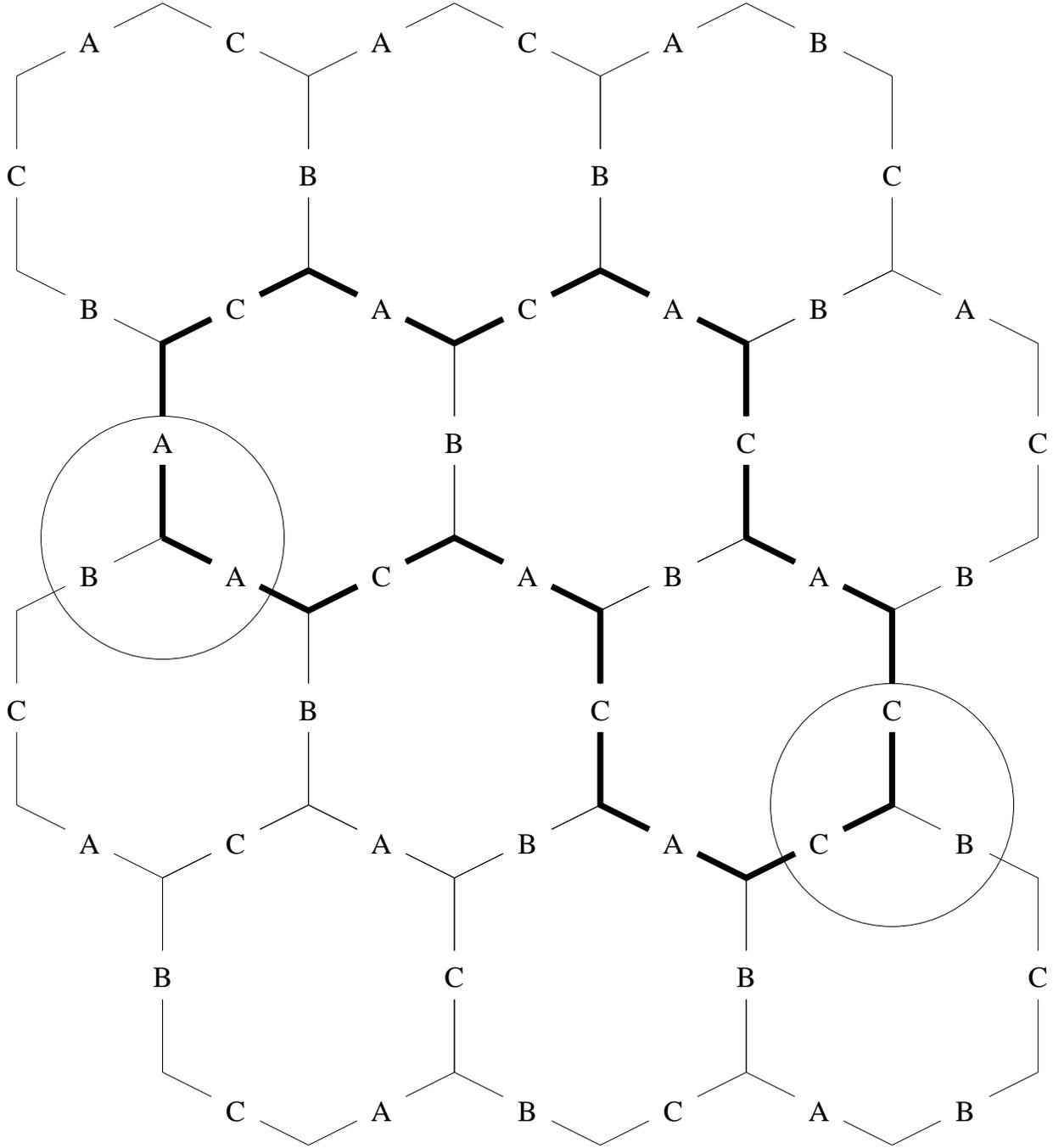
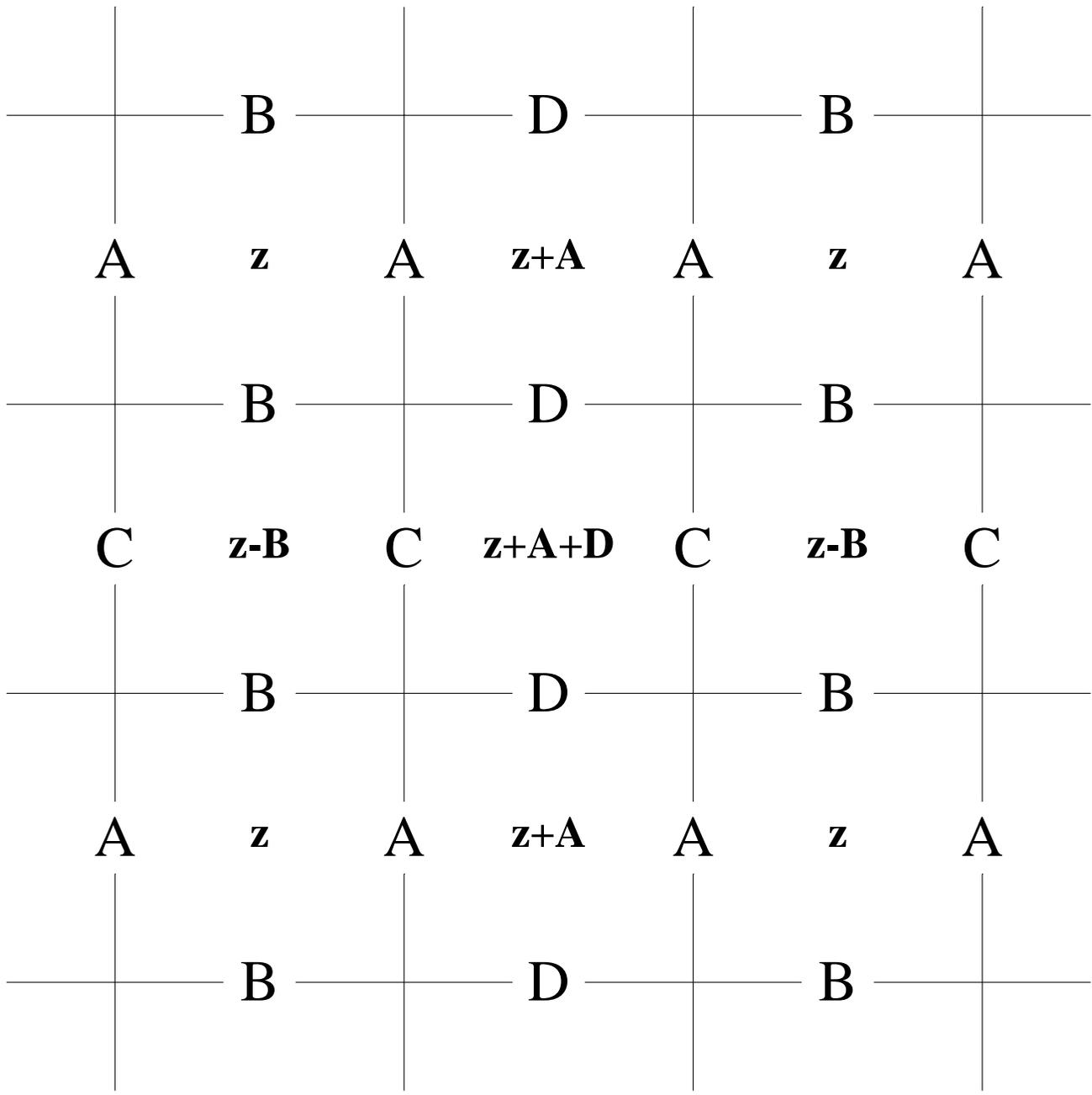
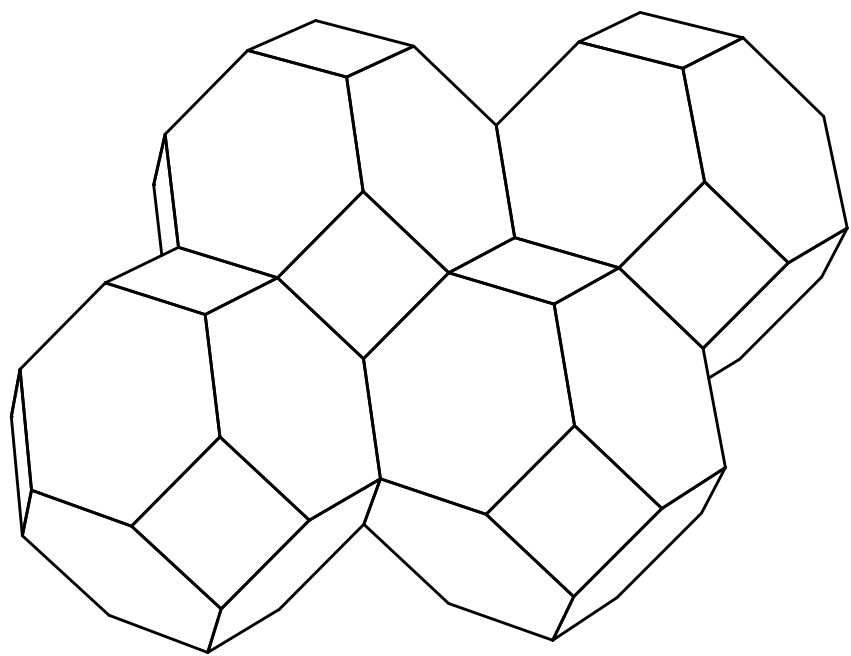
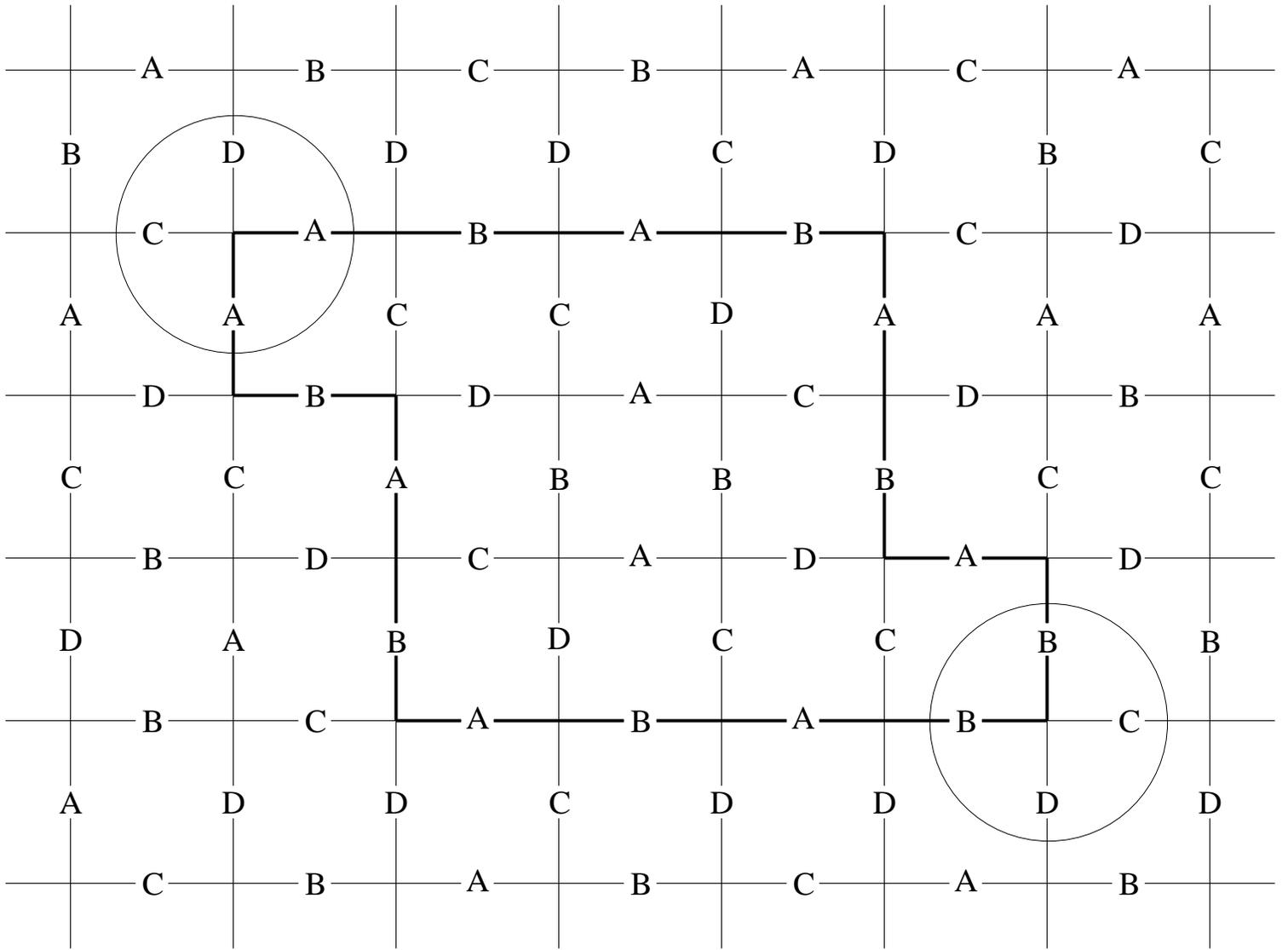
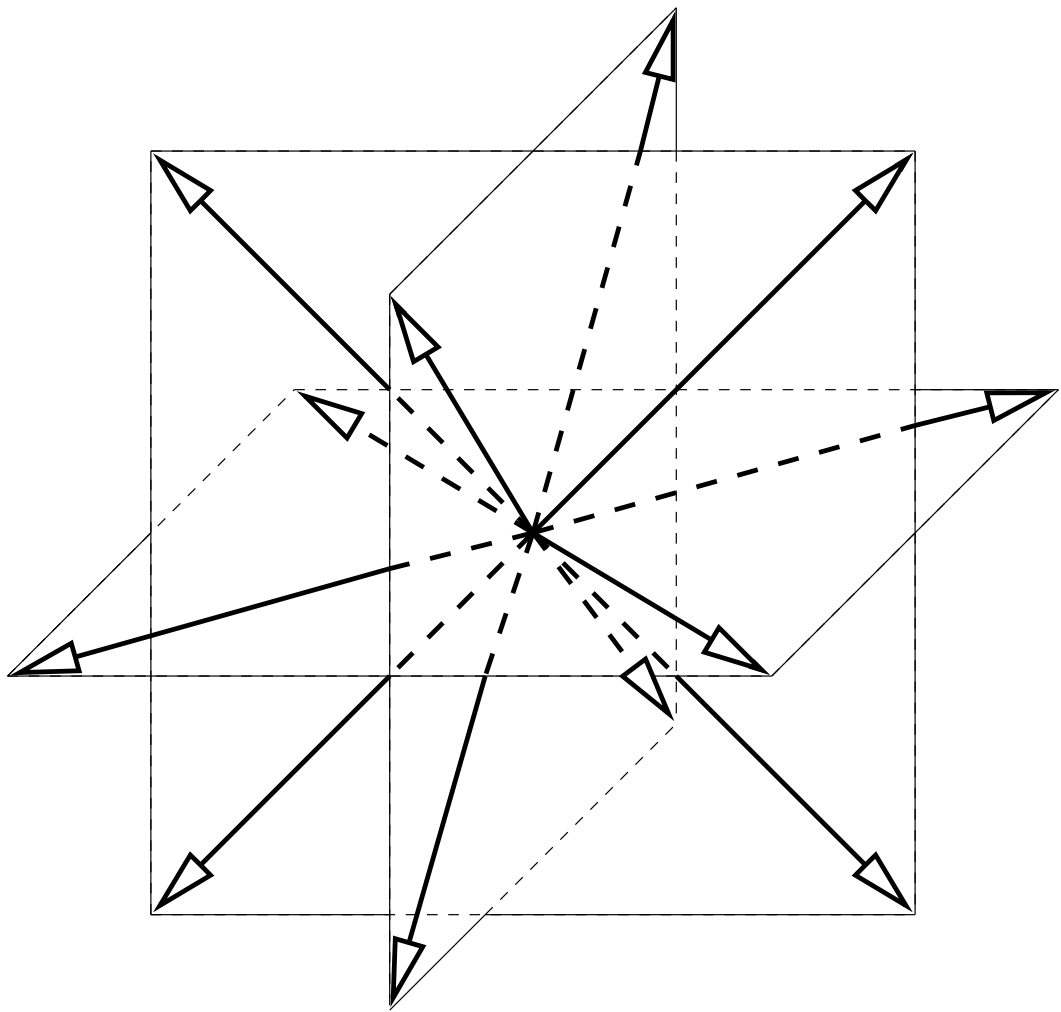
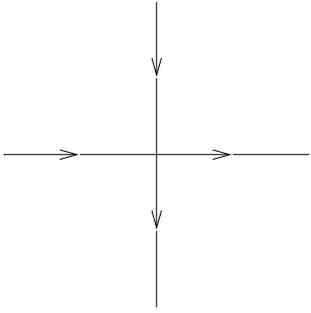
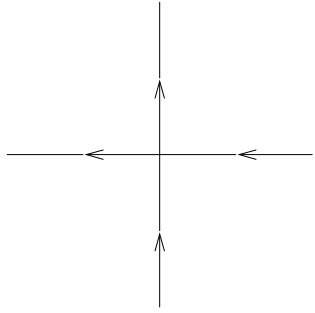
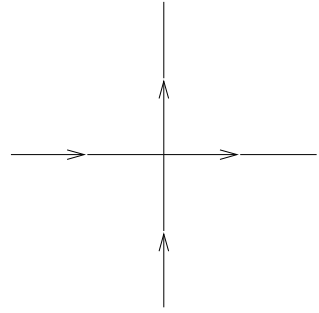
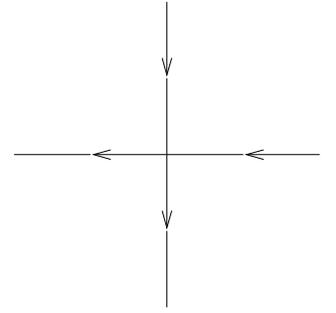
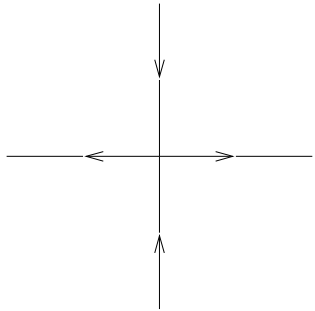
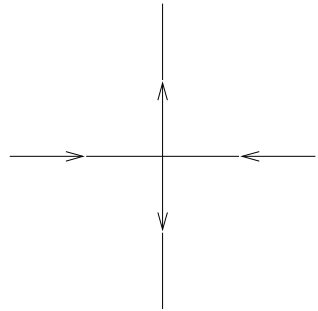
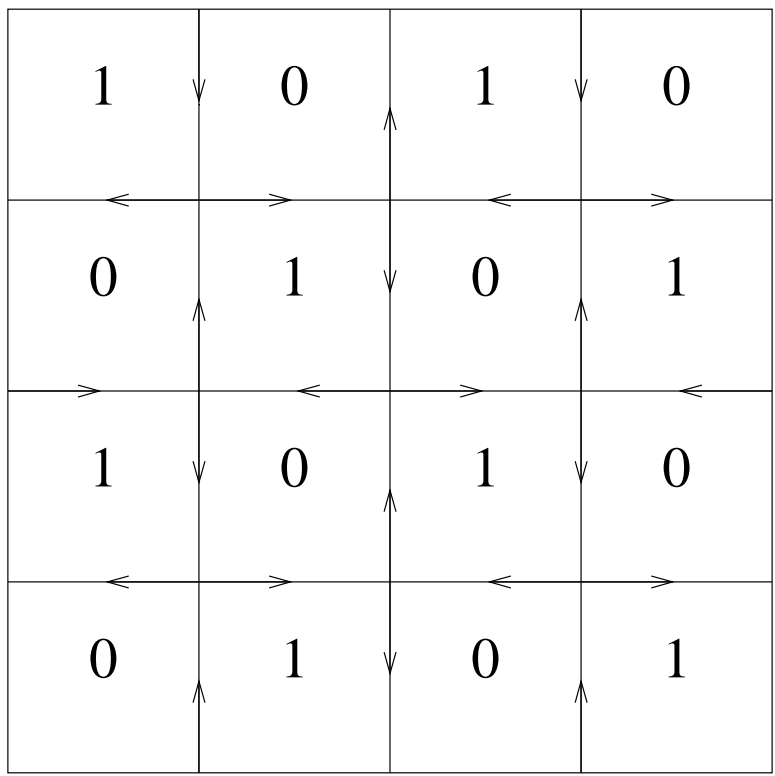
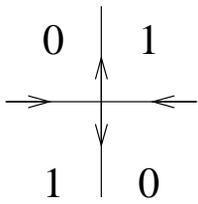
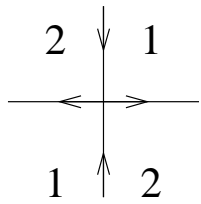
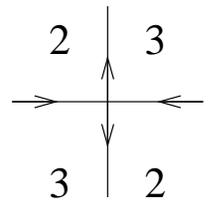
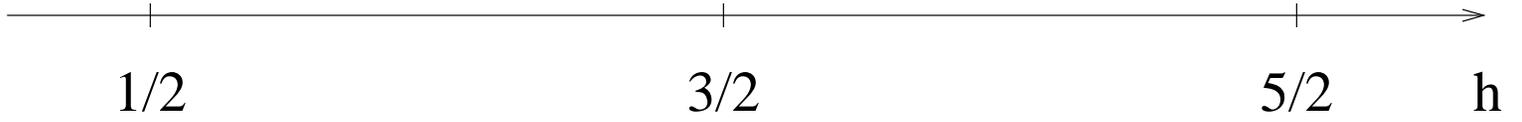
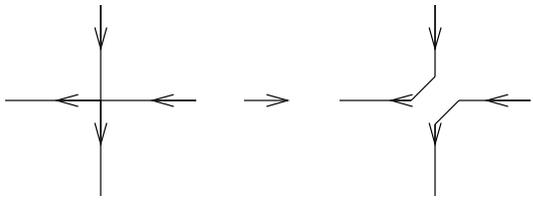
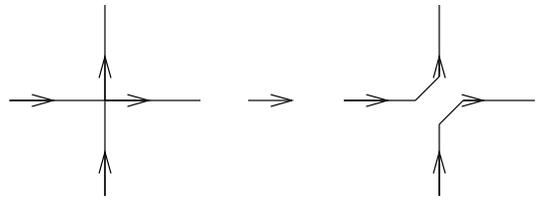
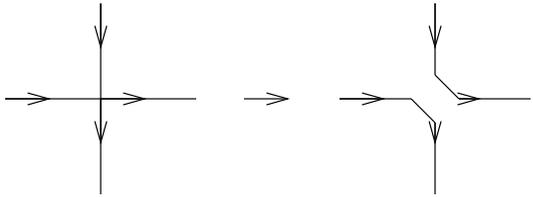
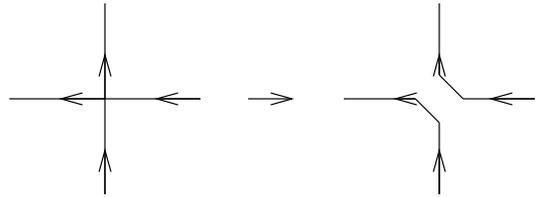
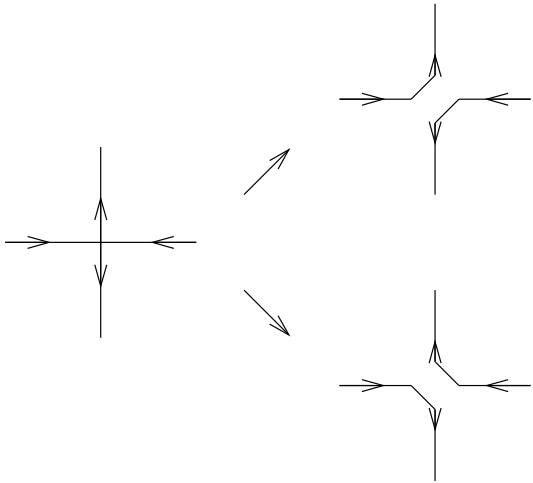
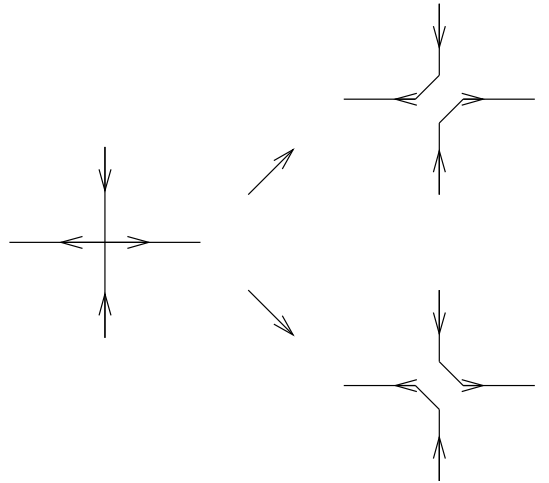
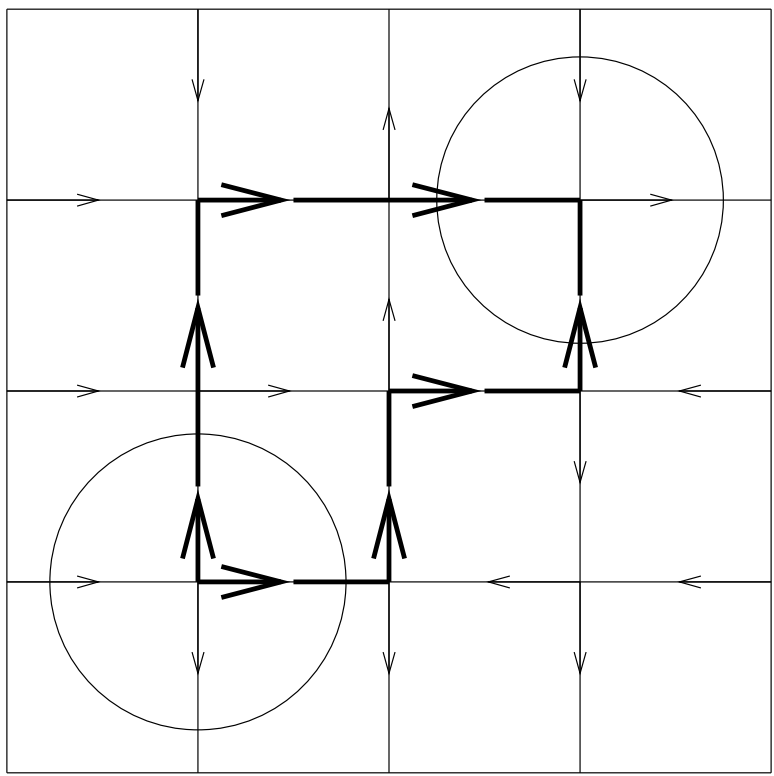
\end{document}